\documentclass[]{amsart}
\usepackage{times}
\usepackage[authoryear]{natbib}
\usepackage[]{graphicx}
\usepackage {amssymb}
\usepackage {amsmath}
\usepackage{pdflscape}
\usepackage{multirow}
\usepackage{booktabs}
\usepackage{xr}

\makeatletter
\def\paragraph{\@startsection{paragraph}{4}%
  \z@\z@{-\fontdimen2\font}%
  {\normalfont\bfseries}}
\makeatother

\usepackage[margin=0.8in]{geometry}

\title{Principled type I error rate inflation in two-arm clinical trial designs with external control information borrowing}
\author{Silvia Calderazzo, Manuel Wiesenfarth, Vivienn Weru and Annette Kopp-Schneider}
\date{\today}

\begin{document}

\label{firstpage}


\begin{abstract}
External information borrowing is often considered in order to improve a clinical trial's efficiency. The Bayesian approach borrows such external information by specifying an informative prior distribution. A potential issue with this procedure is that external and current information may conflict, but such inconsistency may not be predictable a priori. Robust prior choices are typically proposed to limit extreme worsening of operating characteristics (OCs) in these situations. However, trade-offs are still present and in general strict control of type I error (TIE) rate prevents any power gains. In this context, principled justifications for TIE rate inflation can be of interest. We investigate two-arm trials, with a focus on external/historical control information borrowing. We illustrate OCs trade-offs and propose an interpretable approach for external information borrowing. The approach analytically links observed prior-data conflict with allowances for TIE rate inflation and power loss. The approach does not rely on a robust prior specification, but can instead be interpreted as an adaptive choice of Bayes - or, equivalently, frequentist - test decision thresholds under the available informative prior. In addition, it can be used to evaluate any dynamic borrowing approach from a frequentist testing standpoint, and to guarantee robustness with respect to misspecification of the data generating process (i.e., design prior) in Bayesian evaluations. A development for both Normal and binomial outcomes is provided.
\end{abstract}

\maketitle

\section{Introduction}

Borrowing information from external data sources can improve the efficiency of a clinical trial design and help cope with sample size constraints. External information is often incorporated via Bayesian designs with informative prior distributions. The main concern in this context is the potential for heterogeneity between the current and external sources of information, or `prior-data conflict', and its implications on the trial's operating characteristics (OCs). Hypothesis testing is often the main focus of early phase trials considering borrowing of external information. 

In a companion article \citep{cald2024}, we focus on the problem of obtaining a compromise between frequentist (no borrowing) and Bayesian (full borrowing) test decisions, with control of the type I error (TIE) rate at a pre-specified level, in one-arm trials. In such situations, the Bayesian test under a fixed informative prior distribution is effectively a uniformly most powerful (UMP) test at some inflated or deflated TIE rate. Different approaches are thus completely equivalent once recalibrated to have the same TIE rate. The main focus of the current article is two-arm studies for which external information about the control arm is available, referred to in the following as `hybrid-control' designs.
In hybrid-control designs TIE rate depends on the true unknown control parameter value and can increase to one \citep{stall2020}. Moreover, different approaches can lead to the same maximum TIE rate while specifying different rejection regions, as will be shown in this work, raising the question of how to select an approach. From a purely Bayesian perspective, the selection of an appropriate prior distribution, genuinely representing the current knowledge about the parameter, suffices. If fully trusted, such a prior can be used both at the design stage to generate potential data outcomes, and at the analysis stage to make inferences. However, some degree of skepticism about prior assumptions is always present, and it increases if the observed data and prior information show strong heterogeneity.
Inferences can be robustified by so-called dynamic borrowing: if external and current information diverge, external information is gradually discarded. Examples of Bayesian dynamic borrowing approaches are the robust mixture prior \citep[see e.g.][]{schm2014}, the power prior \citep[see e.g.][]{ibra2000}, and the commensurate prior \citep[see e.g.][]{hobb2012}. Frequentist approaches have also been proposed \citep[see][for a review]{tari2025}. 
The variety of robust approaches available and the difficulty in selecting their tuning parameters show, however, that achieving robustness is a non-trivial task. 
On one hand, robustness of the prior often relates to its tail behavior - and tail behavior relates to small probabilities which are notoriously hard to elicit \citep{berg1985}. On  the other hand, clear guidance on criteria for robustness in Bayesian clinical trials has been lacking. 
The recent FDA draft guidance \citep[][]{fda2026} has paved the way for a more widespread adoption of Bayesian methods in drug development and has shown openness towards the use of Bayesian performance metrics when external information is borrowed. 
The metrics proposed in the draft guidance formally and desirably combine the frequentist power function with design prior information and/or costs. Maximum TIE rate and power at specific points of the parameter space do not have great influence if such points have low prior probability and the respective error rates have low relative cost. However, they play a larger role if the design prior - and/or cost - is misspecified. In this sense, a reasonable behavior of frequentist error rates across feasible regions of the parameter space can also enhance robustness of long-run Bayesian performance metrics. Moreover, TIE rates are still widely understood and easily communicated metrics - arguably, mainly due to historical reasons.  

Here, we present an approach which allows combining dynamic borrowing with explicit control of the TIE rate.
In Section \ref{sec:general} we highlight the connections between frequentist and Bayesian hypothesis testing in two-arm hybrid-control studies. 
In Section \ref{sec:compromise} we introduce the proposed compromise approach, while
Section \ref{sec:simulation}  illustrates its behavior in simulations. Section \ref{sec:casestudyMAP} shows its application in an exemplary case study. 
Finally, in Section \ref{sec:conclusion} we  discuss our approach in the broader context of Bayesian OCs, optimality and robustness, and contrast it with alternative solutions. Derivations and applications for Binomial outcomes are provided in the Web Appendix.

\section{Two-arm testing}
\label{sec:general}

\paragraph{Bayes and frequentist test decisions}
\label{sec:pfun}

For illustrative purposes, we focus on a two-arm clinical trial with Normal outcomes, and specifically let $\bar{y}_C \sim N(\theta_C, \sigma/\sqrt{n_C})$ and $\bar{y}_T \sim N(\theta_T, \sigma/\sqrt{n_T})$ be the outcomes of the control and the treatment arm, respectively, where $n_j$, $j=C,T$, is the sample size of each arm, and we assume for simplicity $\sigma$ known. Denote also $y=\{\bar{y}_C,\bar{y}_T\}$. Assume that we are interested in the difference $\delta=\theta_T-\theta_C$, and in particular on testing the null hypothesis $H_0: \delta \leq \delta_0$ versus the alternative hypothesis $H_1: \delta> \delta_0$, where $\delta_0$ is the minimal clinically important difference. We assume that external or historical information is available for the control arm, and the interest is in augmenting the current control arm with such information. 
Let \emph{a priori} $\theta_C \sim \pi_C = N(\mu_{C}, \sigma_{C})$, where $\mu_C$ corresponds to the control arm mean of the external trial and $\sigma_{C}=\sigma/\sqrt{n_{0C}}$, assuming that $n_{0C}$ external samples with the same standard deviation as the current trial ones are available. Let also $\theta_T \propto 1$, i.e., no external information is available for the treatment arm. Denote by $\pi$ the joint prior distribution for $\theta_C$ and $\theta_T$. The posterior distribution of $ \delta | y$ is $N(\mu_{\delta|y}, \sigma_{\delta|y})$, where
 $\mu_{\delta|y}  = \bar{y}_T - \frac{\mu_{C} A_C + \bar{y}_C}{1+A_C}$ and 
$\sigma_{\delta|y}  = \sqrt{\frac{\sigma^2}{n_T} + \frac{\sigma^2/n_C}{1+A_C}}$,
and $A_C=\sigma^2/n_C \sigma^2_{C}$ is the ratio of the data to prior variance in the control arm.

The Bayesian test decision (BD) rejects the null hypothesis if $P(\delta \leq \delta_0 | y) \leq \gamma$, where $\gamma$ is referred to in the following as `test decision threshold' and is an appropriately chosen value, typically 0.025 or 0.05. Therefore, the BD rejects if
\begin{equation}
\frac{\mu_{\delta|y} - \delta_0}{\sigma_{\delta|y}} > z_{1-\gamma},
\label{eq:rej}
\end{equation}
where $z_{1-\gamma}$ denotes the $1-\gamma$ quantile of the standard Normal distribution. Note that, when $A_C=0$, i.e., the control arm prior is flat and no information is borrowed, (\ref{eq:rej}) reduces to the frequentist $z$-test with frequentist test decision (FD) to reject if 
\begin{equation}
Z=\frac{\bar{y}_T - \bar{y}_C  - \delta_0}{\sqrt{\frac{\sigma^2}{n_T} + \frac{\sigma^2}{n_C}}} > z_{1-\kappa},
\label{eq:freq}
\end{equation}
and the type I error rate is $\alpha=\kappa$. Note that, from a decision-theoretic perspective, $\kappa$ can be interpreted as the ratio $c_0/(c_0+c_1)$, where $c_0$ is the cost of making a type II error and $c_1$ as the cost of making a type I error. The frequentist and Bayesian test decisions are therefore optimal under paradigm-specific criteria which depend on such a cost ratio. A full decision-theoretic justification of the BD and FD optimality is provided in Web Appendices \ref{sec:freqOpt} and \ref{sec:BayesOpt}.
Rearranging (\ref{eq:rej}), 
the condition for rejection becomes 
\begin{equation}
\frac{ \bar{y}_T - \theta_{T}- \frac{\bar{y}_C - \theta_{C}}{1+A_C}}{\sqrt{\frac{\sigma^2}{n_T}  + \frac{\sigma^2/n_C}{(1+A_C)^2} }} > \frac{\delta_0  + \frac{\theta_C}{1+A_C} - \theta_T +  \frac{\mu_{C} A_C}{1+A_C} + z_{1-\gamma} \sqrt{\frac{\sigma^2}{n_T} + \frac{\sigma^2/n_C}{1+A_C}}}{\sqrt{\frac{\sigma^2}{n_T}  + \frac{\sigma^2/n_C}{(1+A_C)^2} }}.
\label{eq:twoarm}
\end{equation}
The left-hand side of (\ref{eq:twoarm}) follows a standard Normal distribution. The power function, which describes the probability to reject the null hypothesis, becomes, therefore \citep{stall2020}
\begin{equation*}
\beta^{\pi}(\theta_C, \theta_T) = 1-\Phi \left[\frac{\delta_0 + \frac{\theta_C}{1+A_C} - \theta_T +  \frac{\mu_{C} A_C}{1+A_C} + z_{1-\gamma} \sqrt{\frac{\sigma^2}{n_T} + \frac{\sigma^2/n_C}{1+A_C}}}{\sqrt{\frac{\sigma^2}{n_T}  + \frac{\sigma^2/n_C}{(1+A_C)^2} }} \right],
\end{equation*}
where the notation $\beta^{\pi}$ is used to refer to the power function of the BD under the joint prior $\pi$.
When $\theta_T=\theta_C+\delta_0$, let $\alpha^\pi=\beta^{\pi}(\theta_C, \theta_T=\theta_C+\delta_0)$, i.e.,
the TIE rate at $\delta_0$ of the test incorporating external information. Note that, as shown in Web Appendix \ref{sec:regression}, reparameterization to a regression model with a dummy variable for the treatment effect leads to identical inferences.
As noted in the literature \citep{stall2020}, TIE rate can approach 0 or 1, as $\theta_C \rightarrow -\infty$ or $\theta_C \rightarrow \infty$, respectively. Therefore, TIE rate cannot be uniformly controlled under the BD by increasing the test decision threshold $\kappa$ if $\kappa$ is a fixed value, unless more restrictive assumptions about the mean of the control group $\theta_C$ are made. 
In order to cap TIE rate,  one can adopt a Bayesian dynamic borrowing approach, which would however induce an \emph{a priori} unknown upper bound, or make $\kappa$ dependent on the observed data $\bar{y}_C$, as shown in the following section. 
Note that we have concentrated on a hybrid-control design: when informative priors are available for both the control and the treated arm, and the two arms have the same data to prior variance ratio $A_j$, TIE rate is controlled (see Web Appendix \ref{sec:indpriors}).

\paragraph{Calibration of the Bayes test to achieve a pre-specified TIE rate}
\label{sec:cal}

Suppose we wish to calibrate the Bayesian test so that it achieves a target TIE rate $\alpha^*$ for any $\theta_C$. 
One can aim at matching the FD and the BD for given observed data. In particular, let $\kappa^{BD}$ be a probability threshold such that the BD in (\ref{eq:rej}) is obtained when using the frequentist test in (\ref{eq:freq}) with $\kappa=\kappa^{BD}$. 
To obtain $\kappa^{BD}$, note that (\ref{eq:rej}) is equivalent to
\begin{equation}
\frac{\bar{y}_T - \bar{y}_C - \delta_0}{\sqrt{\frac{\sigma^2}{n_T}  + \frac{\sigma^2}{n_C}}} > \frac{(\mu_C - \bar{y}_C)\frac{ A_C}{1+A_C}  +  z_{1-\gamma} \sqrt{\frac{\sigma^2}{n_T} + \frac{\sigma^2/n_C}{1+A_C}}}{ \sqrt{\frac{\sigma^2}{n_T} + \frac{\sigma^2}{n_C}}} = z_{1-\kappa^{BD}},
\label{eq:equiv}
\end{equation}
which is the frequentist test in (\ref{eq:freq}) with updated $\kappa=\kappa^{BD}$, for fixed $\gamma$ (e.g., the usual 0.025). Note that, on the other hand, $\kappa^{BD}=\kappa^{BD}(\bar{y}_C)$, i.e., it is \emph{not} constant, and depends on the observed control arm mean $\bar{y}_C$. For notation convenience, we will make dependence on $\bar{y}_C$ explicit for test decision thresholds only when they are not part of a subscript. Note that the duality between prior information and frequentist test decisions thresholds was shown in \cite{penn2007} for one-arm studies. We can equivalently derive the data-dependent threshold $\gamma^{FD}(\bar{y}_C)$ providing the FD under the Bayes test. In particular, assume now a fixed $\kappa$ - e.g., 0.025 - for the frequentist test in (\ref{eq:equiv}) which provides the FD. By setting the right-hand side of (\ref{eq:equiv}) to $z_{1-\kappa}$, and inverting with respect to $\gamma$ we obtain
\begin{equation}
    z_{1-\gamma^{FD}}=\left[ \frac{z_{1-\kappa} \sqrt{\frac{\sigma^2}{n_T} + \frac{\sigma^2}{n_C}}   +(\bar{y}_C-\mu_C)\frac{ A_C}{1+A_C} }{\sqrt{\frac{\sigma^2}{n_T} + \frac{\sigma^2/n_C}{1+A_C}}} \right],
\label{eq:realThr}
\end{equation}
which is the critical value to be adopted in (\ref{eq:rej}) to obtain the test decision in (\ref{eq:freq}). As typically $\kappa$ and $\gamma$ are chosen to be standard values such as 0.025 for both the Bayes and the frequentist test, different test decisions are obtained for the same data outcomes under the two approaches. 
A graphical representation of $\kappa^{BD}(\bar{y}_C)$ and $\gamma^{FD}(\bar{y}_C)$ is shown in Figure \ref{fig:decThr}, and a full derivation of (\ref{eq:equiv}) is provided in Web Appendix \ref{sec:calibration}.

If we adopt a data-dependent recalibration approach, we can thus obtain the UMP test at any chosen level $\alpha^*$ under the Bayes test (\ref{eq:rej}). 
However, since the test is UMP, no power gains are possible without TIE rate inflation \citep{kopp2020}. Considerations similar to the one-arm situation apply \citep{kopp2020,cald2024}: prior information supports with higher probability a certain range of parameter values. If the prior is correct, information borrowing will lead to gains. In the hybrid-control design, both TIE rate deflation and power gains can be achieved by improving estimation of $\theta_C$ %
and, in turn, $\delta$, therefore improving test decisions under both the null and alternative hypotheses. In a neighborhood of $\mu_C$, a small conflict $|\mu_C-\theta_C|$ can still lead to gains, due to the reduction in posterior variance for the treatment effect $\delta$. However, if strong conflict is present, TIE rate inflates or power decreases: this happens when the posterior estimate for $\theta_C$ is either strongly biased upwards, leading to 
fewer rejections compared to the FD, or downwards, leading in contrast to 
more rejections than the FD. 
This has been empirically observed in \cite{viele2014}. 
The following sections will focus on a borrowing mechanism that explicitly links allowance for TIE rate inflation with the degree of observed conflict.

\section{Compromise approach}
\label{sec:compromise}

\begin{figure}
   \centering
   \includegraphics[width=1\textwidth]{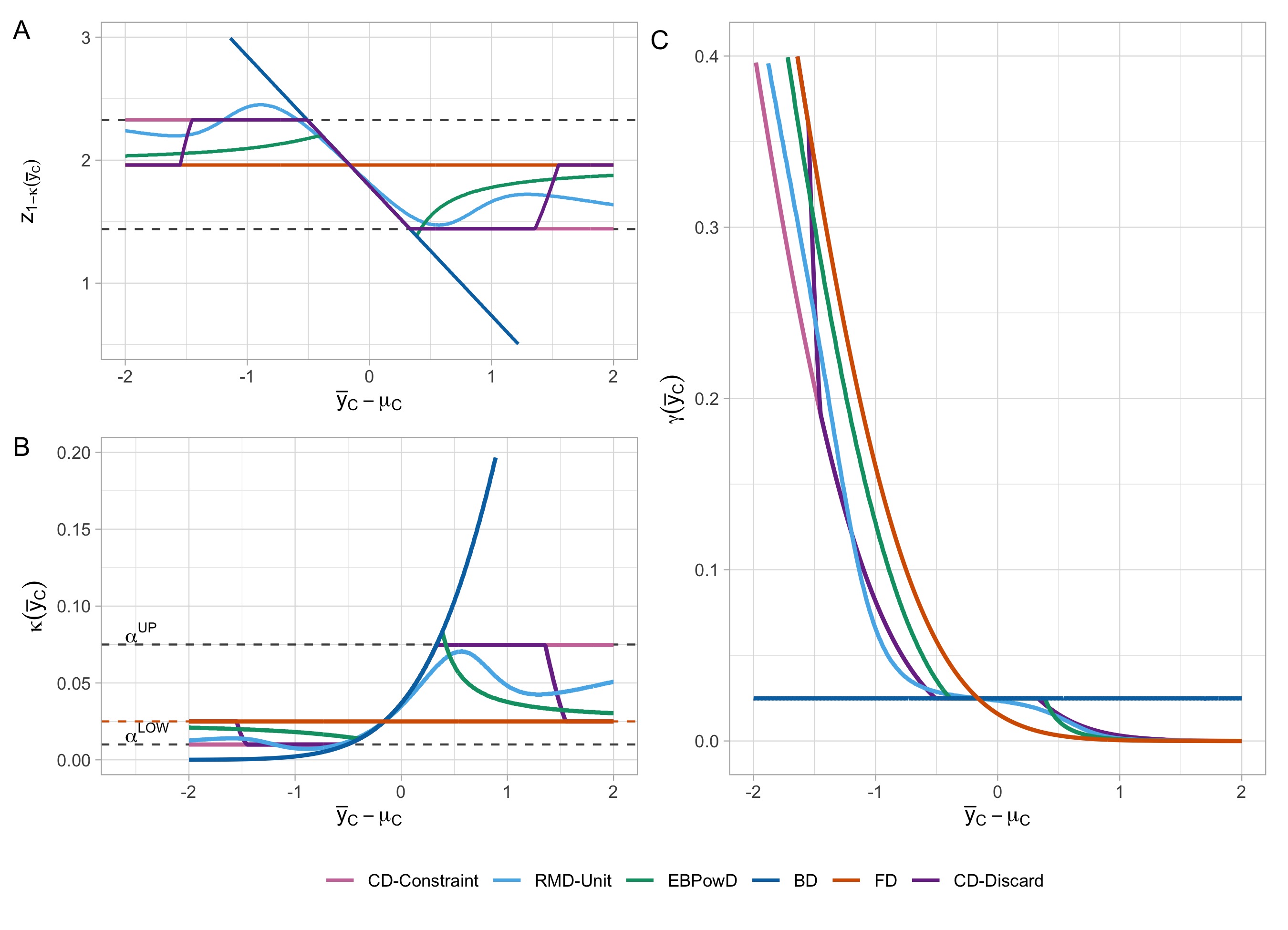}
   \caption{Critical $z$-value (A), frequentist test decision threshold (B), and Bayes test decision threshold (C) for different borrowing approaches as a function of the observed conflict $\bar{y}_C-\mu_C$. Normal outcomes with known variance $\sigma^2=1$, $n_C=n_T=20$. Informative prior $\pi_C=N(0,\sigma_C=\sigma/\sqrt{10})$, $t = 4$, and $p = 4$ for the CDD approach. The dashed horizontal lines in the left plots correspond to the $z$-values and $\kappa(\bar{y}_C)$ bounds for the CD approaches.}
\label{fig:comparison1}
\end{figure}

We extend the methodology in \cite{cald2024} to implement a compromise test decision (CD),
i.e., a test decision which incorporates external information while capping inflation of TIE rate in a pre-specified way. 
We propose two approaches: in the first, full borrowing is allowed only to the extent that TIE rate is not inflated and power deflated above pre-specified values,
and it will be denoted CD-Constraint (CDC); in the second, conflict is accounted for in a more gradual way, with a full reversal to no borrowing when $\bar{y}_C>> \mu_C$ or $\bar{y}_C << \mu_C$, and will be denoted CD-Discard (CDD).
A graphical representation of the frequentist and Bayes test thresholds induced by the CDC and CDD approaches for an exemplary case are shown in Figure \ref{fig:comparison1}, and details are provided in the following. We first focus for simplicity on Normal outcomes, as the power function is available in closed form. The methodology is not however restricted to Normal outcomes: it only requires identification of  test decision thresholds that induce the BD under the frequentist test. Derivations for Binomial outcomes are provided in Web Appendix \ref{sec:binomial}.

\paragraph{Derivation}

We keep the same notation as in Section \ref{sec:general}, and in particular let $\theta_C \sim \pi_C$, and $\theta_T \propto 1$ \emph{a priori}.
Following (\ref{eq:equiv}), recall that for given $\bar{y}_C$ and $\gamma$ under the BD,
\begin{equation}
{\kappa}^{BD}(\bar{y}_C)=1-\Phi \left[\frac{ (\mu_{C}- \bar{y}_C) \left(\frac{A_C}{1+A_C}\right) +  z_{1-\gamma} \sqrt{\frac{\sigma^2}{n_T} + \frac{\sigma^2/n_C}{1+A_C}}}{\sqrt{\frac{\sigma^2}{n_T}  + \frac{\sigma^2}{n_C}}}\right].
\label{eq:testdecH}
\end{equation}
is the test decision threshold required under the frequentist test (\ref{eq:freq}) to obtain the BD.
We then `compromise' $\kappa^{BD}(\bar{y}_C)$ by defining the CDC test threshold as
\begin{equation}
  \kappa^{CDC}(\bar{y}_C) = 
    \max(\min({\kappa}^{BD}(\bar{y}_C), \alpha^{UP}),\alpha^{LOW}), 
\label{rule:borr}
\end{equation}
where $\alpha^{UP}$ and $\alpha^{LOW}$ are pre-specified upper and lower bounds for the TIE rate. Indeed, if $\kappa^{CDC}(\bar{y}_C)$ is \emph{never} above $\alpha^{UP}$ or below $\alpha^{LOW}$, then control of TIE rate at such levels is guaranteed (see Web Appendix \ref{sec:tierate} for an analytical derivation of the resulting TIE rate). The CDC can be seen as an approximation to the solution of a constrained Bayesian optimization problem where the BD, which is optimal under full borrowing, is selected as long as TIE rate constraints are guaranteed. This point is further investigated in Web Appendix \ref{sec:optimality}.  Note that, if $\alpha^{UP}$ is selected equal to $\kappa=\alpha$, the CDC would always induce more conservative test decisions than the FD 
, hence TIE rate and power would be uniformly lower than for FD. This suggests selecting $\alpha^{UP}>\kappa$. While $\alpha^{UP}$ caps TIE rate inflation, $\alpha^{LOW}$  caps the power losses, so power considerations would aid its specification. 

From (\ref{eq:testdecH}), $\kappa^{BD} (\bar{y}_C) > \alpha^{UP}$ iff
$\frac{A_C (\bar{y}_C - \mu_C)}{1+A_C}  > -\left[z_{1-\alpha^{UP}}\sqrt{\frac{\sigma^2}{n_T} + \frac{\sigma^2}{n_C}}- z_{1-\gamma} \sqrt{\frac{\sigma^2}{n_T}+ \frac{\sigma^2/n_C}{1+A_C}} \right] $, and, analogously, $\kappa^{BD} (\bar{y}_C) < \alpha^{LOW}$ iff
$\frac{A_C (\bar{y}_C - \mu_C)}{1+A_C} < -\left[z_{1-\alpha^{LOW}}\sqrt{\frac{\sigma^2}{n_T} + \frac{\sigma^2}{n_C}}-z_{1-\gamma} \sqrt{\frac{\sigma^2}{n_T}+ \frac{\sigma^2/n_C}{1+A_C}}  \right]$.
The CD approach therefore links observed conflict with allowances for TIE rate inflation or deflation. This is possible due to the monotonicity of the distribution of $\bar{y}_C$, as well as TIE rate and power, with respect to $\theta_C$, as shown formally in Web Appendix \ref{sec:monotonicity}. Note that there is a small range of parameter values where ${\kappa}^{BD}(\bar{y}_C) \geq \kappa$ and $\bar{y}_C \leq \mu_C$: in this range the BD induces less conservative test decisions than the FD, but the observed conflict acts in the opposite direction by `pushing' the observed mean upwards.  Indeed, the BD and FD thresholds coincide for 
$\bar{y}_C =\mu_C- z_{1-\gamma} \left[\sqrt{\frac{\sigma^2}{n_T} + \frac{\sigma^2}{n_C}}- \sqrt{\frac{\sigma^2}{n_T}+ \frac{\sigma^2/n_C}{1+A_C}} \right] \frac{1+A_C}{A_C}$,
from which we observe that, for $A_C> 0$, the equality is always satisfied at a point $\bar{y}_C < \mu_C$. From (\ref{eq:testdecH}), we also have that for $\gamma=\kappa$ (e.g., 0.025), ${\kappa}^{BD}(\bar{y}_C=\mu_C) \geq \kappa$. It follows that, if one wishes to employ the BD when no conflict is observed, then $\alpha^{UP} \geq {\kappa}^{BD}(\bar{y}_C=\mu_C)$.

In addition to controlling of TIE rate and power loss, full discard of external information may be desired for very large conflict. A gradual discard can be obtained by adopting 
\begin{eqnarray}
\label{eq:realThr3}
  \kappa^{CDD}(\bar{y}_C) &=& \max\left(\min\left(\left( \frac{\kappa}{\kappa^{BD}(\bar{y}_C)}\right)^w \kappa^{BD}(\bar{y}_C), \alpha^{UP}\right),\alpha^{LOW}\right) \\
\nonumber w &=&  \min\left( \left(\frac{|\bar{y}_C-\mu_C|}{t \sqrt{\sigma^2/n_C + \sigma^2_C}}\right)^p,1 \right)
\end{eqnarray}
where $t>0$ tunes the observed conflict value at which we fully revert to the FD, and $p\geq0$ controls the speed at which such reversal happens. Note indeed that $w=0$ implies the BD rule, and $w=1$ the FD rule. Note, also, that for data outcomes such that ${\kappa}^{BD}(\bar{y}_C) \geq \kappa$ and $\bar{y}_C \leq \mu_C$, we will in general have $w \neq 0$. In this range, indeed, we observe an increase in ${\kappa}^{BD}(\bar{y}_C)$ despite negative conflict: this is where the decrease in posterior variance compensates for the negative conflict (which would make the BD more conservative), and the BD becomes more liberal than the FD. For large enough $p$, however, its effect is practically negligible (see 
 \ref{fig:kappaCD}), and nearly full borrowing is achieved. If monotonicity of $\kappa^{CDD}(\bar{y}_C)$ is desired, $w$ can be fixed to zero in such a range.
The parameter $t$ corresponds to the number of standard deviations away from zero we would allow the observed difference between the current and external mean to be, before considering external data irrelevant. Note that this is equivalent to computing the prior predictive $p$-value \citep{box1980} for the current observed data, i.e., $P(|\bar{Y}^{PP}_C| \geq |\bar{y}_C|)$ where $\bar{Y}^{PP}_C \sim N(\mu_C, \sigma^2/n_C + \sigma^2_C)$ is the prior predictive distribution, and discard external information when such $p$-value falls below $2\Phi(-t)$. Predictive $p$-values have been widely used to detect prior-data conflict and tune borrowing \citep[see e.g.][]{niko2017,kwia2024}, the novelty we introduce is how they are included in the weight specification (\ref{eq:realThr3}) together with $p$. Note, also, that when the prior is not obtained from historical information and/or is non-normal, a normal approximation of the prior is often still a reasonable working assumption. A situation in which a normal approximation would suggest more conflict than practically relevant, is one in which the prior is heavy-tailed, e.g., if it corresponds to a $t$ distribution. In such cases, external information would automatically be discounted for large observed conflict \citep[][]{ohagan2012,weru2024}. We would argue that in such situations the dynamic borrowing approach in (\ref{eq:realThr3}) is neither appropriate nor needed. However, one may still wish to cap TIE rate via (\ref{rule:borr}). We will further discuss and exemplify these aspects in the case study of Section \ref{sec:casestudyMAP}.

\paragraph{Bayesian testing cost-ratio interpretation} The CD rules also have a Bayesian interpretation in terms of type I and type II error cost ratio adaptation. In particular, from (\ref{eq:realThr}), the test decision threshold to be applied under the Bayes test to obtain the CDs (CDC or CDD) is
\begin{equation}
\gamma^{CD} (\bar{y}_C) =1- \Phi \left[ \frac{z_{1- \kappa^{CD}} \sqrt{\frac{\sigma^2}{n_T} + \frac{\sigma^2}{n_C}}   +(\bar{y}_C-\mu_C)\frac{ A_C}{1+A_C} }{\sqrt{\frac{\sigma^2}{n_T} + \frac{\sigma^2/n_C}{1+A_C}}} \right].
\label{eq:realThr2}
\end{equation}
The latter frames the CD approach in terms of a data-adaptive choice of Bayesian test decision thresholds. As these can be interpreted as a function of the ratio of the cost of making a type I error versus a type II error when using prior $\pi$ (see Web Appendix \ref{sec:BayesOpt}), (\ref{eq:realThr2}) defines essentially prior-data conflict dependent costs. In particular, in order to cap TIE rate, the Bayes test cost of a type II error $\gamma^{CD}(\bar{y}_C)$ is increased compared to $\gamma=0.025$ when observed prior-data conflict is such that $\kappa^{CDC}(\bar{y}_C)<\alpha^{LOW}$. Vice versa, when prior-data conflict is such that $\kappa^{CDC}(\bar{y}_C)>\alpha^{UP}$, the Bayes test cost of a TIE, $1-\gamma^{CD}(\bar{y}_C)$, is increased. Costs are further increased or decreased when reversal to no borrowing or reactivity of the approach to small conflict is desired, as in the CDD approach.  
The BD, FD, and CDs frequentist and Bayes test decision thresholds are exemplified in Figure \ref{fig:comparison1}.

\paragraph{Connections with alternative dynamic borrowing approaches}

Any alternative dynamic borrowing mechanism can be related and contrasted with the CD approach by evaluating their induced test decision thresholds. 
For Normal outcomes, this requires uniquely identifying for each $\bar{y}_C$ the minimum $z$-statistics for which the null hypothesis is rejected. 
As the Bayes decision to reject is a monotone function of the posterior probability of the null, the requirement for the existence and uniqueness of $z_{1-\kappa^{\text{approach}}}$ for given $\bar{y}_C$ is that such posterior probability is monotone in $\bar{y}_T$. Provided that prior adaptations do not depend on $\bar{y}_T$, any Normal (mixture) prior would satisfy the requirement, based on monotonicity of the posterior mean with respect to $\bar{y}_C$. When the posterior form is not available analytically,  monotonicity has to be proven numerically. This seems, however, a minimum requirement of any reasonable borrowing mechanism.

For some borrowing mechanisms such as the power prior approach, the test decision threshold $\kappa^{\text{approach}}$ can be obtained analytically. 
The power prior is obtained by raising the external data likelihood to a power parameter $\zeta$ between 0 and 1, where zero induces full discard of historical information, and 1 full borrowing. For Normal outcomes, this leads to a prior of the form $\theta_C \sim \pi^{PP}_C = N(\mu_{C}, \sigma_{C}/\sqrt{\zeta})$, and consequently
\begin{equation}
{\kappa}^{PP}(\bar{y}_C,\zeta)=1-\Phi \left[\frac{ (\mu_C- \bar{y}_C) \left(\frac{A_C \zeta}{1+A_C \zeta}\right) +  z_{1-\gamma} \sqrt{\frac{\sigma^2/n_C}{1+A_C \zeta} + \frac{\sigma^2}{n_T}}}{\sqrt{\frac{\sigma^2}{n_T}  + \frac{\sigma^2}{n_C}}}\right].
\label{eq:ppthr}
\end{equation}

The Empirical Bayes power prior (EBPow) is constructed by dynamically estimating $\zeta$ via maximization of the marginal likelihood  
\citep{grav2017}. The corresponding threshold can then  be derived by substituting the dynamically estimated power parameter $\hat{\zeta}=\zeta(\bar{y}_C)$ into (\ref{eq:ppthr}), and computing the test decision threshold for the EBPow decision (EBPowD) $\kappa^{EBPowD}(\bar{y}_C)= \kappa^{PP}(\bar{y}_C,\hat{\zeta})$. 
The EBPow is not a fixed prior but does not require the choice of any borrowing parameter, and has been shown to achieve reasonable borrowing performance in realistic sample size situations. Moreover, TIE rate is bounded and reverts to the FD as prior-data conflict becomes large.
An alternative common robust borrowing mechanism, which relies on a fixed prior instead, is the robust mixture prior (RM) \citep[e.g.,][]{schm2014}, where the informative prior $\pi_C$ is mixed with a more dispersed robust component to facilitate discarding of prior information for large observed prior-data conflict. Formally, $\pi^{RM}= w \pi_C + (1-w) \pi_{\text{robust}}$
where $w$ is a weight between 0 and 1. Note that, despite the prior being fixed, the weight adapts according to the observed prior-data conflict, making this prior a dynamic borrowing approach. 
Here we focus on the robust mixture prior with unit information robust component centered at the informative prior mean (RM-Unit), a common choice for the hybrid control situation. Here, $\kappa^{RMD-Unit}(\bar{y}_C)$ can be derived numerically by searching for the smallest value of the $z$-statistics at each $\bar{y}_C$ for which we observe a rejection under the RMD-Unit, which we denote $z_{1-\kappa^{RMD-Unit}}$. For this prior choice, TIE rate would reach 1 for large prior-data conflict. Note that full discounting for extreme conflict can be achieved if a heavier tailed $t$ distribution is adopted for the robust component, or if the robust component is located at the current control arm mean \citep{weru2024}.

The critical values of the FD, BD, EBPowD, RMD-Unit and CDC and CDD for an exemplary situation are shown in Figure \ref{fig:comparison1}. The CDD and EBPowD revert back to the FD critical values for large conflict, while the RMD-Unit does not, a property which leads to the increase of TIE rate mentioned above. The BD leads, as expected, to critical values that can even be below 0 for extreme conflict, a property which is clearly undesirable. 
In Figure \ref{fig:comparison1} we also observe how the EBPowD or the RMD-Unit can be interpreted as specific choices of Bayesian test decision thresholds. Note that capping their TIE rates would also be possible, as for the CDs, as exemplified in the case study of Section \ref{sec:casestudyMAP}.
Figure \ref{fig:comparison1} suggests that different borrowing approaches may lead to similar, or even identical, maximum TIE rates, while inducing very different rejection regions. 
This point is further investigated in Web Appendix \ref{sec:optimality} and in Supplementary Figures \ref{fig:ph1Thr} and \ref{fig:ph1Diff}.

\paragraph{Average type I error rate and power}
\label{sec:ir}

Bayesian or `average' OCs for testing have been recently discussed by \citet{best2024} in the form of average type I error rate and power. While frequentist maximum TIE rate can be seen as the most pessimistic estimate of the probability of making a type I error, average OCs allow incorporating realistic assumptions about the parameter of interest in such evaluation. Average metrics are strongly related to frequentist ones - as they can be seen as averages of frequentist test error rates according to a specific prior, called \emph{sampling} or \emph{design} prior. In this sense, we regard frequentist and average quantities as non-competing and complementary. In the hybrid-control situation, average test error rates are obtained by integrating over the informative prior distribution for the control arm mean $\theta_C$. Formally, following \citet{best2024}
\begin{equation}
\bar{\beta}^{\pi^s_C}(\delta, d)=\int_{-\infty}^{\infty} \beta^{\pi^a}(\theta_C,\theta_T=\theta_C+\delta)   \pi^s_C(\theta_C) d\theta_C,
\label{eq:ir}
\end{equation}
where $\pi^s_C$ denotes the sampling prior for $\theta_C$ and $\pi^a_C$ the analysis prior. The difference lies in the fact that the sampling prior represents the assumed `true' prior and is used for the data-generating process, while  the analysis prior is used for fitting the data and obtaining the posterior distribution and test decisions. In this context $\beta^{\pi^a}$ is the power function obtained when decisions $d$ are taken according to the chosen analysis prior or approach. The sampling and analysis prior can coincide, but more generally this construction can be used to evaluate average OCs when the true data-generating process is consistent with the informative prior $\pi_C$, or a collection of potential data-generating process priors, while the analysis is performed under an alternative, possibly robust, $\pi^a$.
When $\delta\leq\delta_0$, average TIE rate is obtained, while $\delta>\delta_0$ gives the average power. 
Note that when rejection regions differ between approaches, their average error rates can also differ.

\section{Simulation study}
\label{sec:simulation}

We study the properties of the proposed methodology in a simulation study. 
We assume the following simulation setup, for given prior mean for the control arm $\mu_C$,
$\theta_C = \mu_C + c$ and $\theta_T = \theta_C + \delta$,
where $c$ is the control arm conflict, and $\delta$ the treatment effect. We assume $n_C=n_T=20$ or $n_C=n_T=100$, to evaluate a low sample size and large sample size situation.  
We assume a flat prior for $\theta_T$, while we consider prior informativeness for $\theta_C$ equivalent to that obtained by $n_C/2$ samples.
The treatment effect $\delta$ under the alternative hypothesis is specified in order to achieve approximately 80\% power under the FD. We set $\alpha^{UP}=0.075$ and $\alpha^{LOW}=0.01$ for the CDs, as exemplary values. We take $p=4$ and $t=4$ for the CDD.  Additional perameter values, and balanced and unbalanced scenarios are investigated in Web Appendix \ref{sec:addsimnormal}. 
We assume $\mu_C=0$, $c \in [-2,2]$, and known $\sigma^2=1$. We compare in this situation test error rates of the BD, the FD, the CDC, the CDD, the RMD-Unit where the prior for the control arm is chosen as the mixture $0.7 N(\mu_{C},\sigma_{C})+ 0.3 N(\mu_{C},\sigma)$, and the EBPowD. Results are based on $2*10^5$ Monte Carlo samples. An analogous simulation study for Binomial outcomes is reported in Web Appendix \ref{sec:addsimbinomial}.

\paragraph{Frequentist TIE rate and power}

Figure \ref{fig:sim1} and Supplementary Figure \ref{fig:sim2} show TIE rate and power for  $n_C=n_T=20$ and $n_C=n_T=100$, respectively. The BD achieves very low TIE rate and power when $\theta_C-\mu_C$ is small, and is conversely very liberal when $\theta_C-\mu_C$ is large, reaching eventually one (not shown). The RMD-Unit, the EBPowD and CDD display similar behaviors in that they discount information when evidence of potential prior-data conflict is observed. Note, however, that when conflict is large, TIE rate of the RMD-Unit starts increasing again, driven by the residual unit-information component \citep{weru2024}.
The choice between the CDC and CDD has a major impact on frequentist OCs only for large conflict. 

Table \ref{tab:recalibration} compares the approaches in terms of realized TIE rate and power at $\mu_C-\theta_C=0$, and the CDs achieve the closest values to the BD. The comparison is however not fair as they also have higher maximum TIE rates for $\mu_C-\theta_C \in [-2,2]$. Following \cite{kopp2024} we thus recalibrate the TIE rate of all approaches to have the same maximum as the CDs within the range of conflict values considered. The CDs reach power differences closer to zero, and for a wider range of conflict values compared to the EBPowD and RMD-Unit, as can be observed in Table \ref{tab:recalibration}, Figure \ref{fig:sim1} and Supplementary Figure \ref{fig:sim2}. We further comment on this behavior in Web Appendix \ref{sec:calibration2borr}.
One interesting observation is that test error rates of the robust approaches are not \textit{necessarily} bounded between the FD and BD ones: for example, if $\delta$ is between approximately 0.25 and 0.4 and $n=20$, the power of the EBPowD under no conflict is slightly higher than for both the FD and the BD (Supplementary Figure \ref{fig:EBpower}). For the CDs and RMD-Unit, on the other hand, we observe slightly lower power for $\delta=0.25$. Note, however, that this is not an issue from a conditional point of view, as the critical values will be bounded by the FD and BD ones, but it may be less intuitive when considering long-run evaluations.

\paragraph{Average TIE rate and power}

Average TIE rate and power allow giving more weight to \textit{a priori} more probable regions of the parameter space. Supplementary Figure \ref{fig:avgOC} shows the average TIE rate and power of the different approaches when $\pi^s(\theta_C)=\pi^a(\theta_C)$, i.e., the sampling and fitting prior coincide for the BD, and for varying $\delta$. Table \ref{tab:avgOC} shows the values of average TIE rate and power at a range of positive $\delta$ values. Consistently with the results of \citet{best2024}, average TIE rate is equal to $\alpha=0.025$ at $\delta=0$ for the FD and the BD, but the BD also achieves higher power for any $\delta>0$. 
The CDC and CDD tend to behave very similarly, because regions in which their OCs significantly differ have very low \textit{a priori} probability. To compare the CDs with the remaining robust approaches, three scenarios are considered: uncalibrated ($\alpha^{LOW}=0.01$  and $\alpha^{UP}=0.075$, as per all evaluations so far), calibrated to RMD-Unit (with weight equal to 0.25, 0.5, and 0.75), and calibrated to EBPowD. In the calibrated cases, $\alpha^{LOW}$ and $\alpha^{UP}$ are set to match maximum and minimum TIE rates of the RMD-Unit and EBPowD across the considered conflict range. 
In general, the CDs tend to achieve better (i.e., closer to the BD) or comparable average test error rates in a wide range of situations for the same maximum conditional TIE rate inflation and power loss, with larger samples sizes improving overall behavior. In Supplementary Table  \ref{tab:avgOC2} and \ref{tab:avgOC3}, we focus on a situation in which the sample size of the external data is  double of the current one, and one in which it is the same as the current but the current sample size of the control arm is half of that of the treated arm, respectively. We observe similar patterns, with the exception of larger average type I error rate and, for small effect size, power, of the EBPowD as compared to the CDs.

\section{Case study}
\label{sec:casestudyMAP}

We exemplify the approach on the case-study in \citet{best2024}, consisting on a  
proof-of-concept study having as primary outcome the change in the disease activity index from baseline. 
The outcome was assumed normally distributed and a meta-analytic predictive (MAP) prior was available for the placebo group based on data from 6 historical studies. A reasonable range for the disease activity score difference was deemed to be between -150 and 50, where lower scores were indicative of stronger improvement.
The MAP prior was further robustified by mixing it with a more dispersed unit-information component centered at the historical data mean, -50 (RMAP). However, the TIE rate of the design is not known \emph{a priori} nor controlled over the whole parameter range. It can therefore be of interest to combine the MAP or RMAP specification with an explicit cap on TIE rate via the CDC. 
For the sake of illustration, we also hypothesize a situation in which only one of the historical studies is available leading to a non-robust Normal prior. In such a situation, it is sensible to apply both the CDC and CDD, and we focus on the CDD.
We take $\alpha^{UP}=0.07$ and $\alpha^{LOW}=0.002$, i.e., the maximum and minimum TIE rate values which are obtained within approximately the 10\% and 90\% quantiles of the MAP as well as the non-robust prior. In principle, such values should be set by discussions between the sponsor and the regulators. 
For the non-robust prior specification, we fully discard prior information when the observed control arm mean is 3 SDs or more away in the tails of the prior predictive distribution ($t=3$), and we choose $p=4$, as a default choice that guarantees fast adaptation and prevents counterintuitive behaviors as discussed in Section \ref{sec:compromise}. Note that $t$, $p$, and $\alpha^{LOW}$ and $\alpha^{LOW}$, all act together as tuning parameters: if $t$ and/or $p$ is very small, the TIE rate caps may never be reached. 

\begin{figure}
   \centering
   \includegraphics[width=0.95\textwidth]{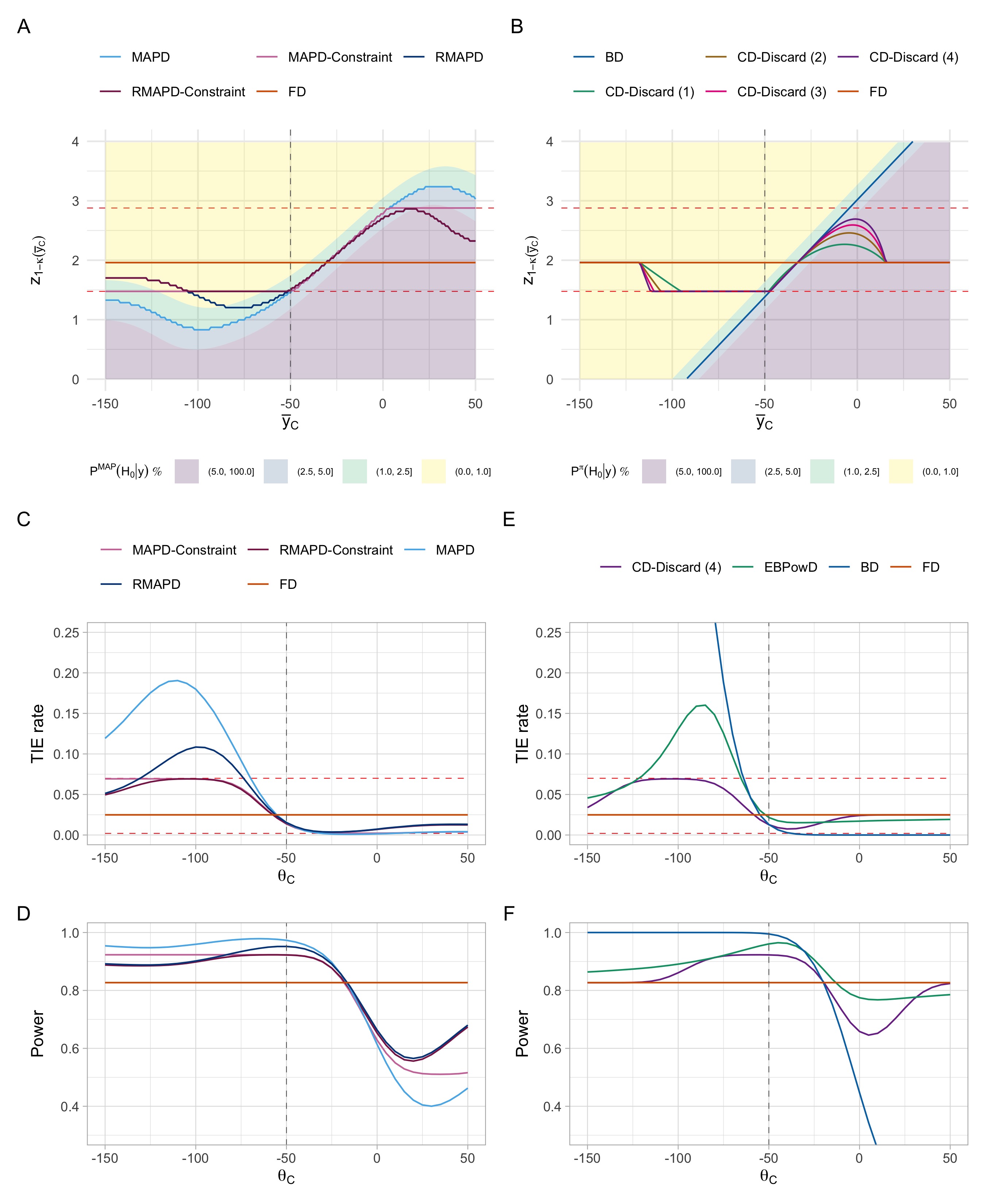}
   \caption{Critical values, TIE rate and Power for different borrowing approaches as a function of true ($\theta_C$) or observed ($\bar{y}_C$) control arm mean for the case study of Section \ref{sec:casestudyMAP}. Normal outcomes, $n_T=40$, $n_C=20$, $\sigma=88$. In panels A, C, and D , a MAP prior is used from the analysis, while in panels B, E, and F, we hypothetically assume that only one historical study is available and the informative prior is normally distributed. The MAP prior was obtained based on 6 historical studies, and was approximated via a mixture of conjugate normal distributions \citep[see][]{rbest}: $\pi^{MAP}(\theta_C | \text{historical data}) = 0.51 N(-51.0, 19.92) + 0.44 N(-46.8, 7.62) + 0.05 N(-54.1, 51.72)$. The prior on the left-hand side corresponds to the first historical study, and is $N(-51,10.23)$.
   The dashed horizontal lines correspond to the TIE rate boundaries for the CD approaches, $\alpha^{LOW}=0.001$ and $\alpha^{UP}=0.15$, $p=4$ and $t=3$ for the CDD (right panels only). The effect of $p$ on the critical values is investigated in panel B.}
\label{fig:case_study_MAP}
\end{figure}

The test decision thresholds, and TIE rate and power for the MAP decisions (MAPD), robust MAP decisions (RMAPD) and non-robust prior (BD) are shown in Figure \ref{fig:case_study_MAP}. Test decisions thresholds for the MAPD and RMAPD are obtained via simulation; however, the final test decisions and OCs computations only requires the evaluation of a frequentist test. 
If we measure the efficiency of the design in terms of power achieved while controlling TIE rate at a pre-specified level, the CDC would in practice always be preferable to the CDD. However, one can argue that extreme observed conflict suggests irrelevance of external information, and thus full discard is sensible, irrespective of power gains. On the other hand, note that, as conflict becomes more extreme, the CDC still reduces the contribution of external information to the analysis: to cap TIE rates when $|\bar{y_C}-\mu_C|$ is very large, external information has to be heavily down-weighted.

A case-study for binomial outcomes is provided in Web Appendix \ref{sec:casestudy}.

\section{Discussion}
\label{sec:conclusion}

In this work, we have explored a rationale for robust external information borrowing in hybrid-control designs. The approach allows explicitly relating observed prior-data conflict to allowances for TIE rate inflation and power loss. 
One could argue that, from a purely Bayesian point of view, the problem can be directly addressed by specifying a prior distribution that genuinely represents the prior beliefs about the parameter of interest. Once this task has been accomplished, optimal Bayesian properties follow (as elaborated in Web Appendix \ref{sec:dectheoretic}), and maximum TIE rate \emph{per se} is no longer a concern. Such a strict Bayesian view relies on two main assumptions: (i) some prior information is available and can be used both at the design and at the analysis stage, and (ii) the prior specification is at least approximately correct, with potential deviations not leading to significant losses. `Correctness' is here defined as a match between the analysis prior and the design prior. The reason point (i) matters is that while vague priors can often be safely used as analysis priors as long as the posterior is proper, using them as design priors is more problematic, and altogether not possible if they are improper. A vague prior, while aimed at representing prior ignorance, actually assigns very large weight to extreme values of the parameter space: for example, when averaging frequentist error rates, results would be mostly driven by values in such extreme regions. 
Point (ii) can be addressed by robust analysis prior specifications.
From a Bayesian perspective, robustness can be investigated in terms of posterior expected loss, or possibly, in situations where repeated use of a procedure is foreseen, integrated risk \citep{berg1985}. In practice, robustness is currently most commonly understood as robustness with respect to maximum TIE rate inflation. When test decisions are taken with respect to a robust informative prior which also caps TIE rate inflation, Bayesian optimality with good frequentist behavior is in principle achieved. The issue is that such a prior is hardly ever available or easy to work with, as discussed in \cite{weru2024}.
In addition, while Bayesian optimality is certainly desirable, and the Bayesian approach philosophically consistent, how to frame such optimality in the context of clinical trial design is an open question. 
The recent FDA draft guidance \citep[][]{fda2026} has provided indications on potential performance metrics, such as the probability of trial success and the probability of a correct decision given a rejection. Additionally, a decision-theoretic approach to hypothesis testing which accounts for the cost of a false negative and a false positive conclusion, was mentioned. 
The aim in the latter is to optimize a (cost-)weighted sum of test error rates, rather than maximize power given a cap on TIE rate \citep[see e.g.][and references therein]{wall2021,cald2020}. 
Weights can follow from elicited costs of type I and type II errors, or from prior probabilities of the null and alternative hypothesis, or from a combination of both, and can also be used to select an optimal $\alpha$ level for a frequentist test. Such a procedure is inherently linked with a posterior expected loss or integrated risk minimization, as the prior probabilities of the null and alternative hypothesis act as a design prior. 
One point to note is that, while all these metrics consistently incorporate costs and/or prior probabilities, they still inherently depend on the power function. Capping error rates globally can be seen as a way to achieve robustness when strong uncertainty about the design prior is present, making its choice less critical.

We have shown that these caps can be implemented by following the BD test decisions threshold for a given range of 'low observed conflict' values of the control arm mean $\bar{y}_C$, and by adopting for larger conflict test decision thresholds equal or more conservative than the FD ones at the selected inflated or deflated TIE rate.
The CDs show favorable frequentist and average OCs. When recalibrating to achieve the same maximum frequentist TIE rate, the power loss of the CDs with respect to the UMP test is the smallest among the considered approaches. Moreover, when looking at average OCs and in the scenarios considered, the CDs tend to reach values closer to the BD than the RMD-Unit, for the same maximum and minimum frequentist TIE rate in almost all scenarios. This is, however, not necessarily always the case, and case-by-case assessments are still recommended. 
 
OCs are evaluated by averaging with respect to the data (frequentist), or with respect to the prior and the data (Bayesian). One could also investigate behavior for realized data outcomes. 
The rationale for such an investigation is that Bayesian procedures are intrinsically conditional, i.e., they aim to optimize the posterior expected loss and perform inference conditioning on the observed data. Web Appendix \ref{sec:dectheoretic} sheds light on why the CDC only approximates an optimal Bayes decision under constraints: additional data outcome pairs leading to a smaller posterior probability of the null under the $\pi$, and therefore a lower posterior expected loss,  
could likely be identified and included in the rejection region without violating the TIE rate constraints 
in the prespecified regions. This may come at a high cost in terms of complexity. A systematic investigation of the potential gains in this context can be of interest, but is left as a potential topic of further research.
It is important to stress that the CDs do not specify a prior. While the induced Bayesian OCs appear sensible in the scenarios considered, and competitive with those of Bayesian approaches with a fixed or empirical Bayes prior, their main aim is to provide a tool to better elicit and interpret underlying assumptions on how trust in external information is translated into test decision rules. For this reason, the approach can be used as a stand-alone method, but also as a way to compare different dynamic borrowing mechanisms and/or further robustify such approaches to potential misspecification of the design prior when considering their long run testing performance.

\bibliographystyle{apalike} \bibliography{biblio2}

\clearpage

\begin{figure}
   \includegraphics[width=\textwidth]{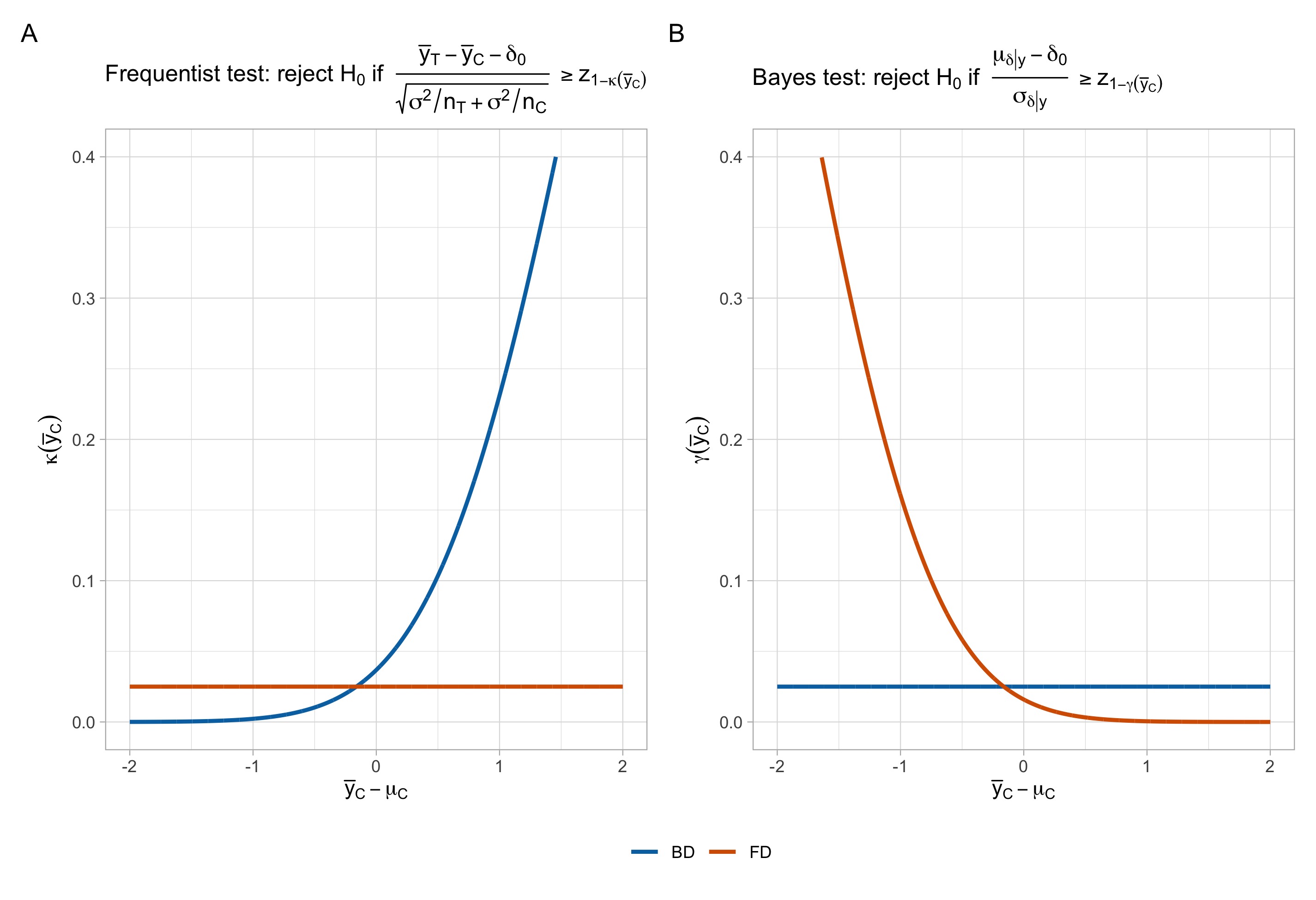}
   \caption{Test decision thresholds under the frequentist (A) and Bayes (B) test for the BD and the FD as a function of the observed conflict $\bar{y}_C-\mu_C$. Normal outcomes with known variance $\sigma^2=1$, $n_C=n_T=20$. Informative prior $\pi_C=N(0,\sigma_C=\sigma/\sqrt{10})$.}
\label{fig:decThr}
\end{figure}

\begin{table}[ht]
\centering 
\begin{tabular}{l|rrrr}
  \hline
\multicolumn{1}{c|}{} & TIE rate & Power & TIE rate Diff. & Power Diff.\\ 
  \hline
  \hline
\multicolumn{5}{l}{$n_C=n_T=20$} \\    
\hline
BD & 1.79 & 88.26 & -7.43 & -91.30 \\ 
  FD & 2.49 & 79.95 & 0.00 & 0.00 \\ 
  RMD-Unit & 1.88 & 86.77 & 0.01 & -0.06 \\ 
  EBPowD & 1.97 & 87.81 & -0.01 & -0.25 \\ 
  CDD & 1.83 & 87.88 & -0.00 & -0.00 \\ 
  CDC & 1.83 & 87.89 & -0.00 & -0.00 \\
   \hline
   \hline
  \multicolumn{5}{l}{$n_C=n_T=100$} \\    
  \hline
  BD & 1.79 & 88.26 & -7.43 & -91.30 \\ 
  FD & 2.49 & 79.95 & 0.00 & 0.00 \\ 
  RMD-Unit & 1.84 & 87.31 & -0.01 & -0.11 \\ 
  EBPowD & 1.97 & 87.81 & -0.00 & -0.24 \\ 
  CDD & 1.83 & 87.88 & -0.00 & -0.00 \\ 
  CDC & 1.83 & 87.89 & -0.00 & -0.00 \\ 
  \hline
\end{tabular}
\label{tab:recalibration}
\caption{TIE rate and power as percent values at $\delta=0.89$ ($n_C=n_T=20$) and  $\delta=0.4$ ($n_C=n_T=100$) for no conflict ($\theta_C=\mu_C$). Difference between the maximum TIE rate and power of each borrowing approach and the FD, when all approaches are calibrated such that their maximum TIE rate over the conflict range [-2,2] is equal to $\alpha^{UP}=0.075$. Normal outcomes with known variance $\sigma^2=1$, $t = 4$, and $p = 4$. Informative prior $\pi_C=N(0,\sigma_C=\sqrt{2\sigma^2/n_C})$, i.e., the external data sample size is assumed to be half of the current control arm one.}
\end{table}

\begin{figure}
   \centering
   \includegraphics[width=1\textwidth]{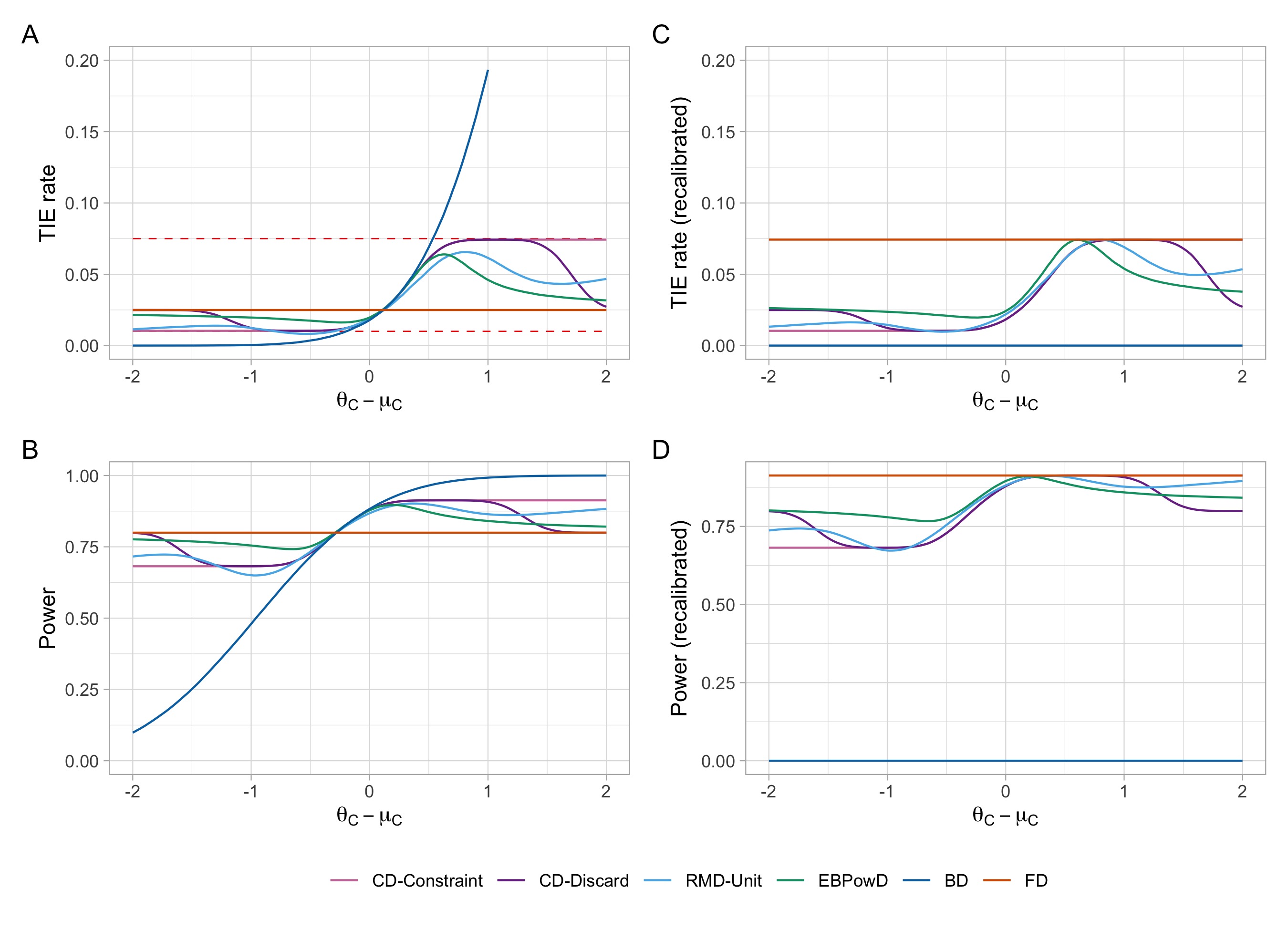}
   \caption{TIE rate (A) and Power (B) and Recalibrated TIE rate (C) and Power (D) for different borrowing approaches as a function of the conflict $\theta_C-\mu_C$. Normal outcomes with known variance $\sigma^2=1$, $n_C=n_T=20$. Informative prior $\pi_C=N(0,\sigma_C=\sigma/\sqrt{10})$. The dashed horizontal lines correspond to the TIE rate boundaries for the CD approaches, $\alpha^{LOW}=0.01$ and $\alpha^{UP}=0.075$, $t = 4$ and $p = 4$ for the CDD.}
\label{fig:sim1}
\end{figure}

\begin{landscape}
\begin{table}
\centering
\begin{tabular}{r|r|rrrrr|rrrrr|}
  \toprule
  \multicolumn{2}{c}{} & \multicolumn{5}{|c|}{$n_C=n_T=20$}  & \multicolumn{5}{c|}{$n_C=n_T=100$} \\      \cmidrule{3-12}
\multicolumn{2}{c}{}  &  \multicolumn{10}{|c|}{$\delta$}  \\ 
\multicolumn{2}{c|}{} & 0 & 0.25 & 0.5 & 0.75 & 1  & 0 & 0.1 & 0.2 & 0.3 & 0.4 \\ \midrule
&BD & 2.45 & 13.68 & 41.11 & 73.79 & 93.35 & 2.45 & 11.78 & 34.12 & 64.29 & 87.29 \\ 
&  FD & 2.48 & 12.17 & 35.32 & 66.03 & 88.44 & 2.48 & 10.57 & 29.34 & 56.52 & 80.74 \\ 
& RMD-Unit  & 2.26 & 12.74 & 38.95 & 71.49 & 92.12 & 2.30 & 11.15 & 32.54 & 62.28 & 85.82 \\ 
 \multirow{2}{*}{\parbox{5cm}{\centering Uncalibrated}} &  CDD   & 2.43 & 13.38 & 40.15 & 72.62 & 92.67 & 2.43 & 11.53 & 33.26 & 63.05 & 86.34 \\   
 & CDC & 2.43 & 13.37 & 40.15 & 72.62 & 92.67 & 2.43 & 11.52 & 33.26 & 63.05 & 86.34 \\    \cmidrule{2-12}
    &  EBPowD & 2.64 & 13.79 & 40.40 & 72.30 & 92.23 & 2.64 & 11.93 & 33.63 & 62.89 & 85.87 \\ 
    \multirow{2}{*}{\parbox{5cm}{\centering Calibrated to EBPowD}} &  CDD  & 2.64 & 13.80 & 40.57 & 72.64 & 92.55 & 2.64 & 11.93 & 33.74 & 63.19 & 86.23 \\ 
  & CDC& 2.64 & 13.80 & 40.57 & 72.64 & 92.55 & 2.64 & 11.93 & 33.74 & 63.19 & 86.24 \\   \cmidrule{2-12}
    &  RMD-Unit (w=0.75) & 2.28 & 12.84 & 39.21 & 71.77 & 92.29 & 2.32 & 11.22 & 32.75 & 62.58 & 86.03 \\ 
 \multirow{2}{*}{\parbox{5cm}{\centering Calibrated to RMD-Unit (w=0.75)}} &  CDD   & 2.33 & 13.02 & 39.56 & 72.14 & 92.46 & 2.36 & 11.32 & 33.02 & 62.88 & 86.32 \\ 
  & CDC  & 2.33 & 13.01 & 39.54 & 72.12 & 92.45 & 2.35 & 11.31 & 33.00 & 62.86 & 86.31 \\        \cmidrule{2-12}
    & RMD-Unit (w=0.5) & 2.23 & 12.47 & 38.01 & 70.31 & 91.46 & 2.27 & 10.86 & 31.82 & 61.20 & 84.99 \\ 
 \multirow{2}{*}{\parbox{5cm}{\centering Calibrated to RMD-Unit (w=0.5)}}   &  CDD & 2.27 & 12.64 & 38.40 & 70.72 & 91.67 & 2.29 & 11.03 & 32.10 & 61.63 & 85.32 \\ 
 & CDC& 2.27 & 12.64 & 38.40 & 70.72 & 91.67 & 2.29 & 11.02 & 32.09 & 61.62 & 85.32 \\     \cmidrule{2-12}
    & RMD-Unit (w=0.25) & 2.24 & 12.21 & 36.96 & 68.85 & 90.54 & 2.25 & 10.60 & 30.76 & 59.57 & 83.58 \\ 
  \multirow{2}{*}{\parbox{5cm}{\centering Calibrated to RMD-Unit (w=0.25)}}  &  CDD  & 2.20 & 12.35 & 37.74 & 70.00 & 91.29 & 2.27 & 10.69 & 30.98 & 59.90 & 83.89 \\
 & CDC  & 2.20 & 12.35 & 37.74 & 70.00 & 91.29 & 2.27 & 10.69 & 30.98 & 59.89 & 83.89 \\ 
       \bottomrule
\end{tabular}
\caption{Average TIE rates (\%) and power (\%) for varying treatment effects $\delta$.  Normal outcomes with known variance $\sigma^2=1$, $n_C=n_T$=20 and $n_C=n_T$=100.  Averages are obtained by integrating over the informative prior for the control arm mean $\pi_C=N(0,\sigma_C=\sigma/\sqrt{n_{0C}})$, corresponding to the analysis prior for the BD. $t = 4$ and $p = 4$ for the CDD. In the uncalibrated scenario, for the CDs $\alpha^{LOW}=0.01$  and $\alpha^{UP}=0.075$, and the weight of the mixture prior $w=0.7$. In each calibrated scenario, $\alpha^{LOW}$ and $\alpha^{UP}$ for the CDC and CDD are set to the minimum and maximum TIE rate observed over the conflict range $[-2,2]$ for the competitor approach.}
\label{tab:avgOC}
\end{table}
\end{landscape}

\newpage

\appendix
\renewcommand{\thesection}{\Alph{section}} 
\setcounter{section}{0}

\section{Decision theoretic aspects}
\label{sec:dectheoretic}

In this section, decision-theoretic justifications for the FD and BD are provided. Moreover, 
we investigate optimality of the CDC in terms of Bayesian constrained optimization. We keep the same notation as the main text, and in particular, let $y$ denote the treatment and control arm outcomes, having probability density function $f(y| \theta_C,\theta_T)$. Assume that we are interested in the difference $\delta=\theta_T-\theta_C$, and in particular on testing the null hypothesis of no (relevant) or negative treatment effect $H_0: \delta \leq \delta_0$ versus the alternative hypothesis $H_1: \delta> \delta_0$. Denote by $\pi$ the joint prior distribution for $\theta_C$ and $\theta_T$.

\subsection{Optimal frequentist decision}
\label{sec:freqOpt}
The frequentist risk related to the testing procedure is obtained by averaging the weighted 0-1 loss function \citep[see also, e.g.,][]{cald2024}
\begin{equation}
    L(\theta,  d_{c_0,c_1})=
    \begin{cases}
    	0 & \text{if} \   d_{c_0,c_1}(y)=I_{\theta_T-\theta_C > \delta_0}\\
      c_0 & \text{if}  \  I_{\theta_T-\theta_C > \delta_0} \   \&  \ d_{c_0,c_1}(y)=0\\
       c_1 & \text{if} \   I_{\theta_T-\theta_C \leq \delta_0} \  \& \  d_{c_0,c_1}(y)=1,
    \end{cases}
    \label{eq:utilityDef}
\end{equation}
with respect to the data distribution, i.e.,
\begin{eqnarray*}
R(\theta_C,\theta_T, d_{c_0,c_1}) &=&\int_Y c_1 I_{d_{c_0,c_1}(y)=1} I_{\theta_T-\theta_C \leq \delta_0} + c_0 I_{d_{c_0,c_1}(y)=0} I_{\theta_T-\theta_C > \delta_0} f(y|\theta_C,\theta_T) dy\\ 
&& c_1 \beta(\theta_C,\theta_T) I_{\theta_T-\theta_C \leq \delta_0} + c_0 [1-\beta(\theta_C,\theta_T)] I_{\theta_T-\theta_C > \delta_0},
\label{eq:fr}
\end{eqnarray*}
where $c_1$ and $c_0$ are the costs associated with making a type I and a type II error, respectively, and $\beta(\theta_C,\theta_T)$ is the power function. The minimum of the maximum risk across the whole two-dimensional parameter space is obtained by taking a decision according the the uniformly most powerful (UMP, frequentist) test, with significance level $\alpha=c_0/(c_0+c_1)$ (the FD). This follows from the fact that for the UMP test, $\beta(\theta_C,\theta_T)$ only depends on, and is monotonically increasing in, $\theta_T-\theta_C$. For the Bayes full borrowing approach (BD), $\beta(\theta_C,\theta_T)$ depends on both $\theta_C$ and $\theta_T$. This leads to local decreases in frequentist risk, and only minor inflation when prior information is consistent with $\theta_C$. Power functions and frequentist risks for $c_0=0.025$, $c_1=0.975$, a grid of $\theta_C$ values, and the borrowing approaches considered in the present article are shown in Supplementary Figures \ref{fig:powFnFreq} and \ref{fig:FRisk}, respectively.

\begin{figure}[h!]
   \centering
   \includegraphics[width=1\textwidth]{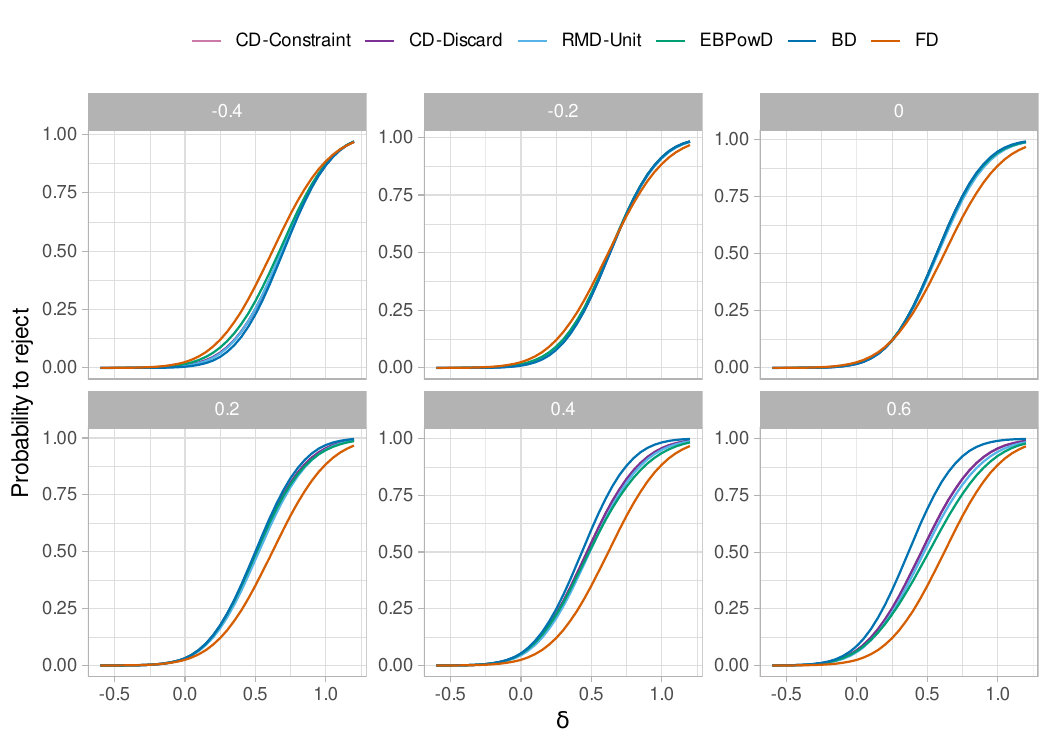}
   \caption{Power functions for a grid of $\theta_C$ values and for different borrowing approaches. Normal outcomes with known variance $\sigma^2=1$, $n_T=n_C=20$. Informative prior $\pi_C=N(0,\sigma_C=\sigma/\sqrt{10})$. when $\theta_C=0$, no conflict is present with the informative prior mean $\mu_C$ and the BD is expected to show the most favorable behavior.  For the CDs, $\alpha^{LOW}=0.01$ and $\alpha^{UP}=0.075$, and $t = 4$ and $p = 4$ for the CDD.}
\label{fig:powFnFreq}
\end{figure}

\begin{figure}[h!]
   \centering
   \includegraphics[width=1\textwidth]{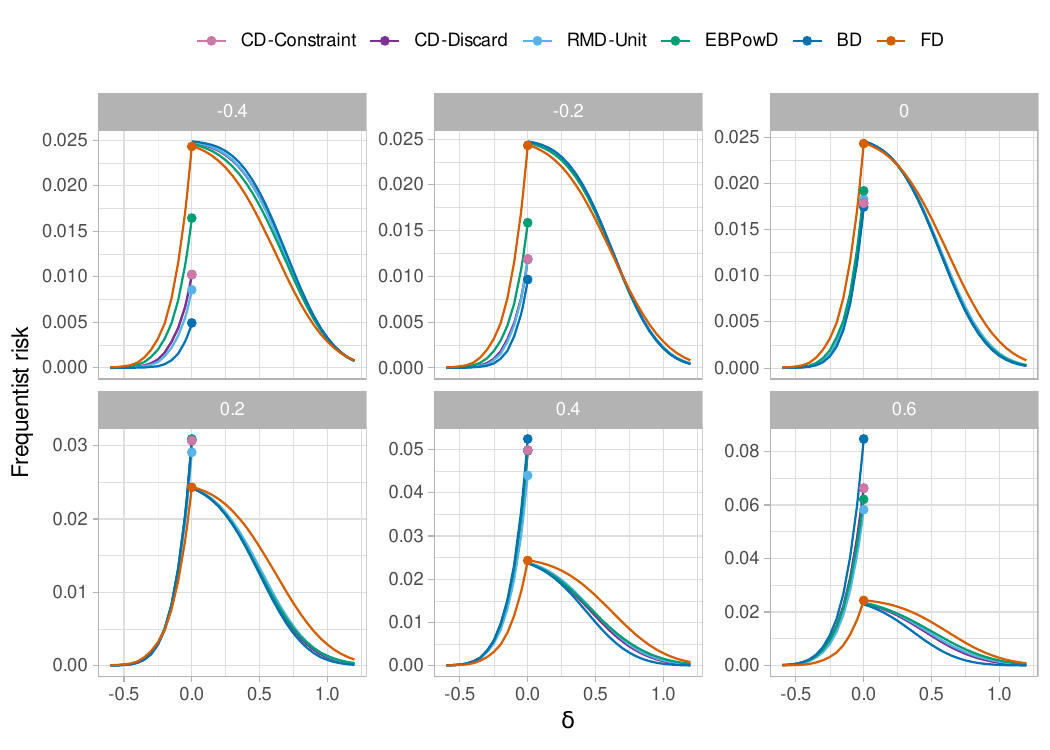}
   \caption{Frequentist risk for a grid of $\theta_C$ values and for different borrowing approaches. Normal outcomes with known variance $\sigma^2=1$, $n_T=n_C=20$. Informative prior $\pi_C=N(0,\sigma_C=\sigma/\sqrt{10})$. when $\theta_C=0$, no conflict is present with the informative prior mean $\mu_C$. $c_0=0.025$, $c_1=0.975$.  For the CDs, $\alpha^{LOW}=0.01$ and $\alpha^{UP}=0.075$,and  $t = 4$ and $p = 4$ for the CDD.}
\label{fig:FRisk}
\end{figure}

\subsection{Optimal Bayes decision}
\label{sec:BayesOpt}
Bayesian optimality is achieved by minimizing the posterior expected loss for each data outcome $y$: for this reason, it is said to optimize the \textit{conditional} behavior of the test procedure. The posterior expected loss is obtained by averaging the weighted 0-1 loss function in (\ref{eq:utilityDef}) with respect to the posterior distribution induced by the prior $\pi$, i.e.,
\begin{align*}
\rho(\pi, d_{c_0,c_1}| y) =&  c_1 P^{\pi}(\theta_T-\theta_C \leq \delta_0 |y) I_{d_{c_0,c_1}(y)=1} + c_0 P^{\pi}(\theta_T-\theta_C > \delta_0|y)  I_{d_{c_0,c_1}(y)=0},
\end{align*}
The posterior expected loss is minimized when the test decision is taken such that $H_0$ is rejected if $P^{\pi}(\theta_T-\theta_C\leq 0 | y) \leq \kappa=c_0/(c_0+c_1)$. 

\subsection{CDC optimality assessment}
\label{sec:optimality}

\begin{figure}
   \centering
   \includegraphics[width=1\textwidth]{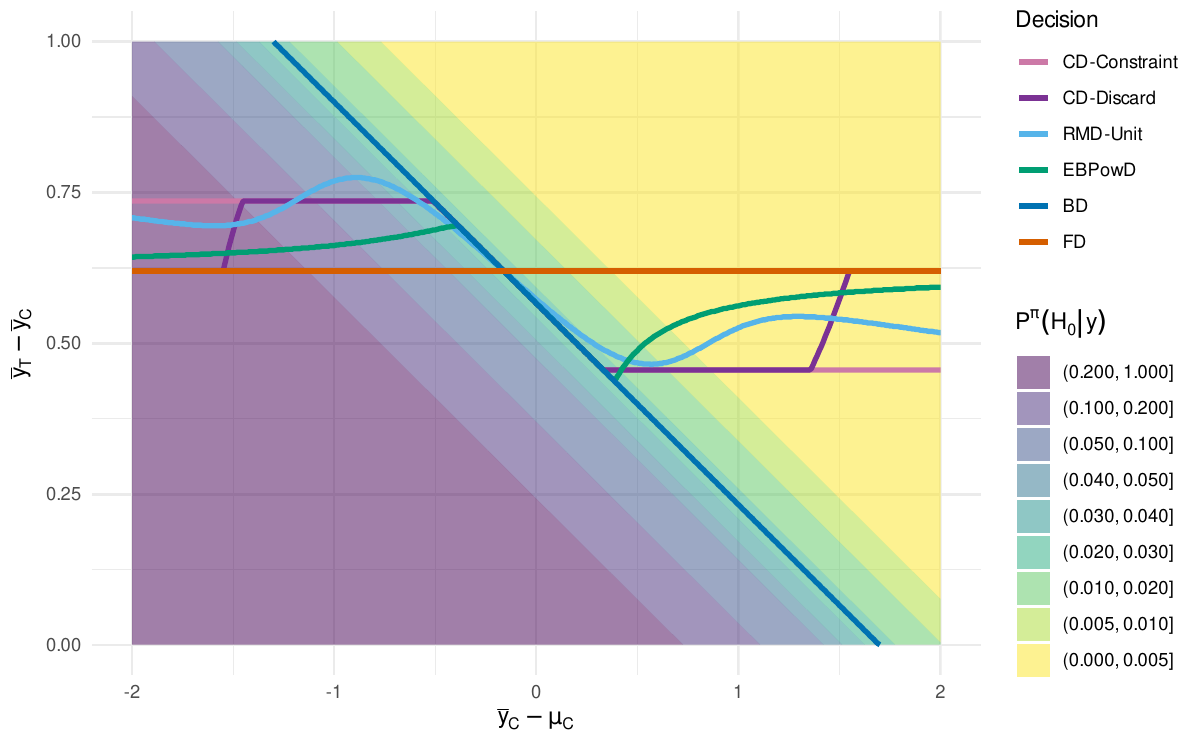}
   \caption{Posterior probability of the null under $\pi_C=N(0,\sigma_C=\sigma/\sqrt{10})$ for varying observed conflict $\bar{y}_C-\mu_C$ and unstandardized effect $\bar{y}_T-\bar{y}_C$ (background). Normal outcomes with known variance $\sigma^2=1$, $n_T=n_C=20$.  For the CDs, $\alpha^{LOW}=0.01$ and $\alpha^{UP}=0.075$, and $t = 4$ and $p = 4$ for the CDD. Boundaries of the rejection region for each borrowing approach are overlaid, the null is rejected for values above the lines.}
\label{fig:ph1Thr}
\end{figure}

In the hybrid-control trial situation, no prior information is available for $\theta_T$. If one would be willing to construct an informative sampling prior $\pi^s_T$ for $\theta_T$, the full Bayesian integrated risk may be computed for any borrowing approach by averaging $R(\theta_C,\theta_T,d_{c_0,c_1})$ with respect to both $\pi^s_C$ and $\pi^s_T$, and approaches could be ranked according to such a value. We do not pursue this approach further as no self-evident prior for $\theta_T$ can be constructed in our set-up, and an improper prior leads the integrated risk to approach zero. Although absence of a proper prior on $\theta_T$ prevents meaningful (absolute) integrated risk comparisons, Bayesian optimality of the procedures can still be compared in terms of posterior expected loss.
Supplementary Figure \ref{fig:ph1Thr} shows the posterior probability of the null, as well as the rejection regions identified by each approach. Notice the similarity to Figure \ref{fig:decThr}. The figure gives insights on why the CDC only approximates an optimal Bayes decision under constraints: pairs with low posterior probability of the null under $\pi$ could likely be identified and incorporated in the rejection region of the CDC test approach, leading to the same TIE rate constraints. 
Differences in posterior probability of the null under $\pi$ and a vague prior $\pi_0$ for pairs giving non-concordant test decisions are shown in Supplementary Figure \ref{fig:ph1Diff}.

\begin{figure}
   \centering
   \includegraphics[width=1\textwidth]{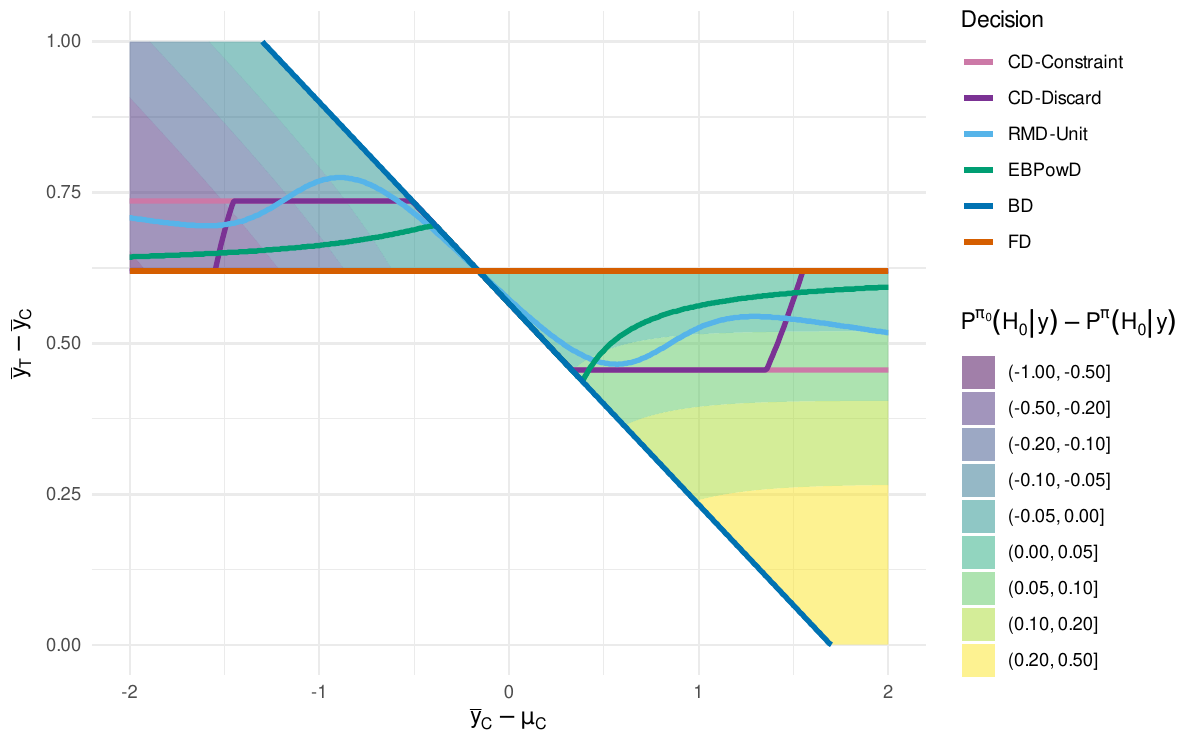}
   \caption{Difference in posterior probabilities of the null under $\pi_C=N(0,\sigma_C=\sigma/\sqrt{10})$ (inducing the BD), and under a vague prior $\pi_0 \propto 1$ (inducing the FD) for the control arm mean, for varying observed conflict $\bar{y}_C-\mu_C$ and unstandardized effect $\bar{y}_T-\bar{y}_C$ (background). Normal outcomes with known variance $\sigma^2=1$, $n_T=n_C=20$.  For the CDs $\alpha^{LOW}=0.01$ and $\alpha^{UP}=0.075$, $t = 4$ and $p = 4$ for the CDD. Boundaries of the rejection region for each borrowing approach are overlaid. Only regions for which the BD and FD test decision differ are colored.}
\label{fig:ph1Diff}
\end{figure}

\section{Derivations for the case of independent informative priors for both the control and treated arm}
\label{sec:indpriors}

In this section we report the expressions for the quantities in Section \ref{sec:general} of the main text for the more general situation in which informative independent priors are available for both the control and the treated arm. We show how informativeness of each arm's prior influences point-wise power gains under the BD after recalibration of the FD's TIE rate, as well as the impact of parametrization when considering a regression model instead. Finally, we outline monotonicity properties of the test induced by the BD, which serve as a theoretical justification for the testing procedure implied by the CDs. 

\subsection{Power function} 

Let \emph{a priori} $\theta_C \sim \pi_C = N(\mu_{C}, \sigma_{C})$ and $\theta_T \sim  \pi_T = N(\mu_{T}, \sigma_{T})$. Also, let $\pi$ be the bivariate Normal prior distribution for $\theta_C$ and $\theta_T$. The null and alternative hypotheses and the data outcomes distributions are as specified in the main text. The posterior distribution of $ \delta | y$ is $N(\mu_{\delta|y}, \sigma_{\delta|y})$, where
\begin{align*}
 \mu_{\delta|y} & = \frac{\mu_{T} A_T + \bar{y}_T}{1+A_T} - \frac{\mu_{C} A_C + \bar{y}_C}{1+A_C}\\
\sigma_{\delta|y} & = \sqrt{\frac{\sigma^2/n_T}{1+A_T} + \frac{\sigma^2/n_C}{1+A_C}},
\end{align*}
and $A_C=\sigma^2/n_C \sigma^2_{C}$, $A_T=\sigma^2/n_T \sigma^2_{T}$ are the ratios of the data to prior variance in each arm.

As outlined in the main text, the BD rejects the null hypothesis if
\begin{equation}
\frac{\mu_{\delta|y} - \delta_0}{\sigma_{\delta|y}} > z_{1-\gamma}.
\label{eq:rej2}
\end{equation}
Note that the frequentist test with TIE rate is $\alpha=\kappa$ is obtained in this case when $A_T=A_C=0$.
To obtain the power function for the BD in the general case, we follow the same steps as in the main text, and in particular from (\ref{eq:rej2}) 
the condition for rejection becomes 
\begin{equation}
\frac{ \frac{\bar{y}_T - \theta_{T}}{1+A_T} - \frac{\bar{y}_C - \theta_{C}}{1+A_C}}{\sqrt{\frac{\sigma^2/n_T}{(1+A_T)^2}  + \frac{\sigma^2/n_C}{(1+A_C)^2} }} > \frac{\delta_0 -\frac{ \mu_{T}  A_T}{1+A_T} +  \frac{\mu_{C} A_C}{1+A_C}  + \frac{\theta_C}{1+A_C} - \frac{\theta_T}{1+A_T} + z_{1-\gamma} \sigma_{\delta|y}}{\sqrt{\frac{\sigma^2/n_T}{(1+A_T)^2}  + \frac{\sigma^2/n_C}{(1+A_C)^2} }}.
\label{eq:twoarm2}
\end{equation}
Once again, the left-hand side of (\ref{eq:twoarm2}) follows a standard Normal distribution. The probability to reject function becomes, therefore
\begin{equation}
\beta^{\pi}(\theta_C, \theta_T) = 1-\Phi \left[\frac{\delta_0 + \frac{\theta_C}{1+A_C} - \frac{\theta_T}{1+A_T} - \frac{ \mu_{T}  A_T}{1+A_T} +  \frac{\mu_{C} A_C}{1+A_C} + z_{1-\gamma} \sigma_{\delta|y}}{\sqrt{\frac{\sigma^2/n_T}{(1+A_T)^2}  + \frac{\sigma^2/n_C}{(1+A_C)^2} }} \right],
\label{eq:prej2}
\end{equation}
When $\theta_T=\theta_C+\delta_0$, we define $\alpha^\pi=\beta^{\pi}(\theta_C, \theta_T=\theta_C+\delta_0)$.
Inspecting (\ref{eq:prej2}), we notice that the dependence on $\theta_C(=\theta_T-\delta_0)$ is no longer present if $1/(1+A_T) = 1/(1+A_C)$, i.e., if the data to prior variance ratio is the same in each arm. Let $A$ be the common variance ratio. In such a situation, a UMP test at level $\alpha^\pi=1-\Phi \left[\frac{A(\mu_C-\mu_T+\delta_0) +\delta_0}{\sqrt{\sigma^2/n_T + \sigma^2/n_C}}  + z_{1-\gamma} \sqrt{A+1} \right]$ is obtained.

\subsection{Point-wise calibration of the frequentist test to the BD TIE rate}

While TIE rate cannot typically be capped when informative priors are adopted, we can investigate the power difference between the borrowing approach (i.e., the analysis under the informative prior distribution $\pi$) and the approach calibrated to the TIE rate of the BD \emph{at each given unknown $\theta_C$}. The power function under the frequentist approach can be obtained by letting $\sigma_{C} \rightarrow \infty$ and $\sigma_{T} \rightarrow \infty$ and assuming $\gamma=\kappa$. It reads
\begin{equation}
\beta^{\text{frequentist}}(\theta_C, \theta_T) = 1-\Phi \left[ \frac{\delta_0 + \theta_C - \theta_T}{{\sqrt{\frac{\sigma^2}{n_T} + \frac{\sigma^2}{n_C}}}} + z_{1-\kappa} \right].
\label{eq:prejfr}
\end{equation}
To obtain the calibrated power function, we replace $z_{1-\kappa}$ with $z_{1-\alpha^{\pi}}=\Phi^{-1}\left[1-\alpha^{\pi}\right]$ in (\ref{eq:prejfr}): Under the null hypothesis, the same TIE rate is thus obtained under the calibrated and the borrowing approach. The calibrated power function reads
\begin{equation*}
\beta^{\text{calibrated}}(\theta_C, \theta_T) = 1-\Phi \left[\frac{\delta_0 + \theta_C - \theta_T}{\sqrt{\frac{\sigma^2}{n_T} + \frac{\sigma^2}{n_C}}} + \frac{\delta_0   + \theta_C \left(\frac{1}{1+A_C} - \frac{1}{1+A_T}\right) - \frac{ \mu_{T}  A_T}{1+A_T} +  \frac{\mu_{C} A_C}{1+A_C} + z_{1-\kappa} \sigma_{\delta|y}}{\sqrt{\left(\frac{\sigma/\sqrt{n_T}}{1+A_T}\right)^2  + \left(\frac{\sigma/\sqrt{n_C}}{1+A_C} \right)^2}} \right].
\label{eq:prejfr1}
\end{equation*}
We now wish to compare the power of the full borrowing (BD) $\beta^{\pi}$ and of the calibrated approach $\beta^{\text{calibrated}}$, in particular, we want to evaluate if $\beta^{\text{calibrated}}>\beta^{\pi}$, $\beta^{\text{calibrated}}=\beta^{\pi}$, or $\beta^{\text{calibrated}}<\beta^{\pi}$.
As both power functions are of the form $1-\Phi(x)$, i.e., monotone transformations of their argument $x$, it is enough to compare their arguments to conclude on differences in power.
In particular, with algebraic steps we have
\begin{equation*}
\Phi^{-1}[1-\beta^{\text{calibrated}}]-\Phi^{-1}[1-\beta^{\pi}]=(\theta_T - \theta_C) \left[ \frac{1}{\sqrt{\frac{\sigma^2}{n_T}+ \frac{\sigma^2}{n_C} \frac{(1+A_T)^2}{(1+A_C)^2}}} - \frac{1}{\sqrt{\frac{\sigma^2}{n_T} + \frac{\sigma^2}{n_C} }} \right] +\frac{\delta_0}{\sqrt{\frac{\sigma^2}{n_T} + \frac{\sigma^2}{n_C} }}.
\end{equation*}
It follows that the power difference depends on whether $(1+A_T)^2/(1+A_C)^2$ is greater, equal, or smaller than 1. In particular, if $\theta_T > \theta_C + \delta_0$ and $\delta_0=0$,
\begin{equation}
    \begin{cases}
    	 \beta^{\text{calibrated}}(\theta_C, \theta_T) > \beta^{\pi}(\theta_C, \theta_T) & \text{if} \  \  n_T \sigma^2_{T} < n_C \sigma^2_{C}\\
      \beta^{\text{calibrated}}(\theta_C, \theta_T) = \beta^{\pi}(\theta_C, \theta_T) & \text{if}  \ \  n_T \sigma^2_{T} = n_C \sigma^2_{C}\\
      \beta^{\text{calibrated}}(\theta_C, \theta_T) < \beta^{\pi}(\theta_C, \theta_T) & \text{if} \ \  n_T \sigma^2_{T} > n_C \sigma^2_{C}.
    \end{cases}
    \label{eq:powdiff}
\end{equation}
For $n_C=n_T$, i.e., an equal number of samples in each arm, power gains under borrowing are therefore locally possible if $\sigma^2_{C} < \sigma^2_{T}$ and $\theta_T > \theta_C + \delta_0$. However, we should stress that (\ref{eq:powdiff}) would hold under any value of $\theta_C$, which is in practice unknown: The maximum TIE rate would still reach 1.

\subsection{Regression parametrization}
\label{sec:regression}

An equivalent formulation of the problem with priors specified for the control arm and the treatment effect, can be obtained via reparametrisation of $[\theta_C,\theta_T]^T$ into $[\theta_C, \delta]^T$: this corresponds to a regression setup that adopts dummy coding for the treatment effect.  Here we show that a regression model formulation leads to the same posterior densities for the treatment effect $\delta$, so such a reparametrisation would also not cap TIE rate. Let
\begin{equation*}
\begin{bmatrix}
    \theta_C \\
    \delta
  \end{bmatrix}
  \sim MVN  \left( \mu_0=
\begin{bmatrix}
     \mu_C \\
    \mu_T -\mu_C
  \end{bmatrix} , \sigma^2 \Lambda_0^{-1} =
  \begin{bmatrix}
     1/n_{0C} & -1/n_{0C} \\
    -1/n_{0C} & 1/n_{0C} + 1/n_{0T}
  \end{bmatrix}
  \right),
\end{equation*}
where $n_{0C}=\sigma^2/\sigma_C$, and $n_{0T}=\sigma^2/\sigma_T$. The reparametrized prior follows from $\theta_C \sim N(\mu_C, \sigma^2_C)$ independent of $\theta_T \sim N(\mu_T, \sigma^2_T)$, and $\delta=\theta_T-\theta_C$. Let 
\begin{equation*}
\begin{bmatrix}
    y_C \\
    y_T
  \end{bmatrix}
  \sim MVN  \left( X \beta, \sigma^2 I_{n_C+n_T} \right),
\end{equation*}
where $y_C$ and $y_T$ are the $n_C$ and $n_T$ vectors of observations,$\beta=[\theta_C, \delta]^T$, and $X$ is the design matrix with dummy coding for the treatment effect. The posterior is \citep{ohag2004}
\begin{equation*}
\begin{bmatrix}
    \theta_C | y_C,y_T  \\
    \delta | y_C,y_T 
  \end{bmatrix}
  \sim MVN  \left(\Lambda_{post}^{-1} \left(X^TX\hat{\beta} +\Lambda_0 \mu_0 \right) , \sigma^2 \Lambda_{post}^{-1} \right),
\end{equation*}
where $\Lambda_{post}=\left(X^TX +\Lambda_0 \right)$. Calculations give
\begin{equation*}
\Lambda_{post}^{-1}=
  \begin{bmatrix}
     \frac{1}{n_{0C}+n_C} & - \frac{1}{n_{0C}+n_C} \\
    - \frac{1}{n_{0C}+n_C} & \frac{1}{n_{0T}+n_T} + \frac{1}{n_0C+n_C}
  \end{bmatrix},
\end{equation*} and 
\begin{equation*}
\Lambda_{post}^{-1} \left(X^TX\hat{\beta} +\Lambda_0 \mu_0 \right)=
  \begin{bmatrix}
     \frac{\mu_C n_{0C}+\bar{y}_C n_C}{n_{0C}+n_C}  \\
    \frac{\mu_T n_{0T}+\bar{y}_T n_T}{n_{0T}+n_T} -  \frac{\mu_Cn_{0C}+\bar{y}_Cn_C}{n_{0C}+n_C}
  \end{bmatrix}.
\end{equation*}
The marginal posterior distribution for $\delta$ is $N\left(\frac{\mu_T n_{0T}+\bar{y}_T n_T}{n_{0T}+n_T} -  \frac{\mu_C n_{0C}+\bar{y}_C n_C}{n_{0C}+n_C},\frac{\sigma^2/n_T}{n_{0T}/n_T+1} + \frac{\sigma^2/n_C}{n_{0C}/n_C+1} \right)$, which has mean and variance identical to that obtained by independent updates of the control and treatment arm, and the difference between the resulting posterior distributions. Note, also, that identical results are obtained when $n_{0T}=0$ whether $\theta_C$ and $\delta$ are assumed to be \textit{a priori} correlated or independent (i.e., if the off-diagonal elements of $\Lambda_0^{-1}$ are taken equal to 0). Therefore, when $\sigma_T=\infty$ (the hybrid-control case), results are invariant to the parametrization.

\subsection{Calibration of the Bayes test to a a frequentist test}
\label{sec:calibration}

The Bayes test condition for rejection (\ref{eq:rej2}) is
\begin{align}
\nonumber &\frac{\frac{\mu_{T} A_T + \bar{y}_T}{1+A_T} - \frac{\mu_{C} A_C + \bar{y}_C}{1+A_C} - \delta_0}{ \sqrt{\frac{\sigma^2/n_T}{1+A_T} + \frac{\sigma^2/n_C}{1+A_C}}} > z_{1-\gamma} \\
\nonumber \frac{\bar{y}_T}{1+A_T} - \frac{\bar{y}_C}{1+A_C} &> \delta_0 + z_{1-\gamma} \sqrt{\frac{\sigma^2/n_T}{1+A_T} + \frac{\sigma^2/n_C}{1+A_C}} - \frac{\mu_T A_T}{1+A_T} + \frac{\mu_C A_C}{1+A_C}\\
\nonumber \frac{\bar{y}_T}{1+A_T} - \frac{\bar{y}_C}{1+A_T} &> \delta_0 + z_{1-\gamma} \sqrt{\frac{\sigma^2/n_T}{1+A_T} + \frac{\sigma^2/n_C}{1+A_C}} - \frac{\mu_T A_T}{1+A_T} + \frac{\mu_C A_C}{1+A_C} + \bar{y}_C \left( \frac{1}{1+A_C}- \frac{1}{1+A_T} \right)\\
\label{eq:equiv1}  \frac{\bar{y}_T- \bar{y}_C -\delta_0}{\sqrt{\frac{\sigma^2}{n_T} + \frac{\sigma^2}{n_C}}} &> \frac{ \delta_0 \frac{A_T}{1+A_T} + z_{1-\gamma} \sqrt{\frac{\sigma^2/n_T}{1+A_T} + \frac{\sigma^2/n_C}{1+A_C}} - \frac{\mu_T A_T}{1+A_T} + \frac{\mu_C A_C}{1+A_C} + \bar{y}_C \left( \frac{1}{1+A_C}- \frac{1}{1+A_T} \right)}{\frac{1}{1+A_T} \sqrt{\frac{\sigma^2}{n_T} + \frac{\sigma^2}{n_C}}},
\end{align}
from which we can observe that the frequentist test with test decision threshold
\begin{equation*}
\kappa^{BD} (\bar{y}_C)= 1 - \Phi \left[ \frac{ \delta_0 \frac{A_T}{1+A_T} + z_{1-\gamma} \sqrt{\frac{\sigma^2/n_T}{1+A_T} + \frac{\sigma^2/n_C}{1+A_C}} - \frac{\mu_T A_T}{1+A_T} + \frac{\mu_C A_C}{1+A_C} + \bar{y}_C \left( \frac{1}{1+A_C}- \frac{1}{1+A_T} \right)}{\frac{1}{1+A_T} \sqrt{\frac{\sigma^2}{n_T} + \frac{\sigma^2}{n_C}}} \right]
\end{equation*}
is equivalent to the Bayes test with test decision threshold $\gamma$.
Moreover, denote the right-hand side of (\ref{eq:equiv1}) $z_{1-\alpha^*}$. Rearranging, we obtain
\begin{equation*}
\gamma=1- \Phi \left[\frac{ \frac{z_{1-\alpha^*} \sqrt{\frac{\sigma^2}{n_T} + \frac{\sigma^2}{n_C}}}{1+A_T}  -\delta_0 \frac{A_T}{1+A_T} + \frac{ \mu_{T}  A_T}{1+A_T} - \frac{\mu_{C} A_C}{1+A_C} - \bar{y}_C \left(\frac{1}{1+A_C} - \frac{1}{1+A_T} \right) }{\sqrt{\frac{\sigma^2/n_T}{1+A_T} + \frac{\sigma^2/n_C}{1+A_C}}} \right].
\end{equation*}

\subsection{Monotonicity properties of the test induced by the BD}
\label{sec:monotonicity}

Equations (\ref{eq:prej2}) and (\ref{eq:equiv1}) allow obtaining additional insights into the properties of the test induced by the BD. In particular, from (\ref{eq:prej2}) we can see that: i) TIE rate is monotonically increasing in $\theta_C$ for $A_C>A_T$, and monotonically decreasing for $A_C<A_T$; ii) power is monotonically increasing in $\theta_C$ for given $\delta>\delta_0$, for $A_C>A_T$, and monotonically decreasing for $A_C<A_T$. Finally, from (\ref{eq:equiv1}), the test decision threshold $\kappa^{BD}(\bar{y}_C)$ is monotonically increasing in $\bar{y}_C$ for $A_C>A_T$, and monotonically decreasing for $A_C<A_T$. Therefore, for a given $z$-statistics, the decision to reject under the BD becomes more likely as $\bar{y}_C$ increases when $A_C>A_T$, which is the case in a hybrid-control trial. As the distribution of $\bar{y}_C$ is stochastically increasing in $\theta_C$, the chances to reject the null also increase in $\theta_C$, which is reflected in an increase of the TIE rate. As the `true' distribution of $\bar{y}_C$, the power for given $\delta$, and the TIE rate are all monotonically increasing in $\theta_C$, values of $\bar{y}_C$ can be used as indicators of prior-data conflict, as well as to derive conditions for TIE inflation and power loss constraint. 

\section{Type I error rate of the CDC}
\label{sec:tierate}

The type I error rate of the CDC is obtained as 
\begin{equation*}
\alpha^{CDC}=E_{y|\theta_T=\theta_C+\delta_0} \left[Z=\frac{\bar{y}_T - \bar{y}_C  - \delta_0}{\sqrt{\frac{\sigma^2}{n_T} + \frac{\sigma^2}{n_C}}} > z_{1-\kappa^{CDC}} \right],
\label{eq:TIECD}
\end{equation*}
where $y=[\bar{y}_C,\bar{y}_T]$ and $\kappa^{CDC}$ is defined as in the main text. From the law of total expectation, it follows that
\begin{eqnarray*}
\alpha^{CDC} &=& P \left( Z> z_{1-\kappa^{BD}}| \kappa^{BD}(\bar{y}_C) \in [\alpha^{LOW},\alpha^{UP}] \right) P \left( \kappa^{BD}(\bar{y}_C) \in [\alpha^{LOW},\alpha^{UP}] \right) +\\
&& P \left( Z> z_{1-\alpha^{UP}}| \kappa^{BD}(\bar{y}_C) > \alpha^{UP} \right) P \left(\kappa^{BD}(\bar{y}_C) > \alpha^{UP} \right)+\\
&& P \left( Z> z_{1-\alpha^{LOW}}| \kappa^{BD}(\bar{y}_C) < \alpha^{LOW} \right) P \left(\kappa^{BD}(\bar{y}_C) < \alpha^{LOW} \right)\\
&=& P \left( Z> z_{1-\kappa^{BD}} | \bar{y}_C \in [\bar{y}_C^{LOW},\bar{y}_C^{UP}] \right) P \left( \bar{y}_C \in [\bar{y}_C^{LOW},\bar{y}_C^{UP} ] \right) +\\
&& P \left( Z> z_{1-\alpha^{UP}}| \bar{y}_C < \bar{y}_C^{LOW} \right) P \left(\bar{y}_C < \bar{y}_C^{LOW} \right)+\\
&& P \left( Z> z_{1-\alpha^{LOW}}| \bar{y}_C > \bar{y}_C^{UP} \right) P \left(\bar{y}_C > \bar{y}_C^{UP} \right)
\label{eq:TIECD_der}
\end{eqnarray*}
where $\bar{y}_C^{LOW}$ and $\bar{y}_C^{UP}$ are defined as (see main text)
\begin{eqnarray*}
\label{eq:ycthr1}  \bar{y}_C^{UP} &=& \mu_C-\left[z_{1-\alpha^{UP}}\sqrt{\frac{\sigma^2}{n_T} + \frac{\sigma^2}{n_C}}-z_{1-\gamma} \sqrt{\frac{\sigma^2}{n_T}+ \frac{\sigma^2/n_C}{1+A_C}} \right] \frac{1+A_C}{A_C} \\
\label{eq:ycthr2} \bar{y}_C^{LOW} &=& \mu_C-\left[z_{1-\alpha^{LOW}}\sqrt{\frac{\sigma^2}{n_T} + \frac{\sigma^2}{n_C}}-z_{1-\gamma} \sqrt{\frac{\sigma^2}{n_T}+ \frac{\sigma^2/n_C}{1+A_C}}  \right] \frac{1+A_C}{A_C},
\end{eqnarray*}
and the probability is computed with respect to $f(y|\theta_C,\theta_T=\theta_C+\delta_0)$, i.e. the data distribution under $H_0$.
The first addend on the right-hand side employs the BD test rule, which applies when  $\kappa^{BD}(\bar{y}_C) \in [\alpha^{LOW},\alpha^{UP}]$. If such an interval is extended to the whole parameter space by taking $\alpha^{LOW}=0$ and $\alpha^{UP}=1$, the CDC rule coincides with the BD rule and therefore achieves the same type I error rate as the BD at any $\theta_C$. The next two addends define regions in which a frequentist test at  level either $\alpha^{UP}$ or $\alpha^{LOW}$ is used.
As $\kappa^{BD}(\bar{y}_C)$ depends on $\bar{y}_C$ and $\bar{y}_C$ has non-zero probability of falling in each of the (mutually exclusive) regions (if $\alpha^{LOW}>0$ and $\alpha^{UP}<1$), no rule will exclusively dominate at any $\theta_C$. However, different $\theta_C$ values imply different probabilities for each of the regions. If the probability of a specific region is `large' at a given $\theta_C$, it will mostly determine the resulting TIE rate at that value.

\section{CD for Binomial outcomes}
\label{sec:binomial}

The CDs can be adapted to Binomial outcomes. Let again $\pi_C$ be the prior distribution for $\theta_C \sim Beta(a_C,b_C)$ and $\pi_T$ the prior for $\theta_T \sim Beta(a_T,b_T)$, where $\theta_C$ and $\theta_T$ denote now the true control and treatment arm probability of success, respectively. Lack of prior information on the treatment arm is summarized by setting $a_T=0.5$ and $b_T=0.5$ for reasons which will be explained soon. Let $\pi$ denote the bivariate prior distribution, obtained as the product of the two independent priors $\pi_C$ and $\pi_T$. Due to conjugacy, after observing $y_C$ successes out of $n_C$ patients in the control arm, and $y_T$ successes out of $n_T$ patients in the treatment arm, the posterior distributions are $Beta(a_C+y_C,b_C+n_C-y_C)$ and  $Beta(a_T+y_T,b_T+n_T-y_T)$, respectively. The distribution for the treatment effect $\theta_T-\theta_C$ is obtained from the difference of such posterior distributions, and the test decision is then obtained analogously to the Normal outcome case according to a pre-specified posterior probability threshold $\kappa$. Connections between frequentist and Bayesian testing in such a situation have been investigated in the literature \citep{alth1969,howa1998}. In particular, define as $\pi^0$ a prior which gives the same test decision as the frequentist approach; here we consider $\pi^{0,F}$ for Fisher's exact test and $\pi^{0,U}$ for the unconditional test. Then, $P^{\pi^{0,F}}[\theta_T-\theta_C \leq 0| y_C, y_T]$ is equal to the $p$-value of Fisher's exact test under the joint prior $\pi^{0,F}$ which takes $a_C=1$, $b_C=0$, $a_T=0$, and $b_T=1$ \citep{alth1969,howa1998}. Analogously, $P^{\pi^{F}}[\theta_T-\theta_C \leq 0| y_C, y_T]$ with $a_C=y_{0C}+1$, $b_C=n_{0C}-y_{0C}$, $a_T=0$, and $b_T=1$ is the $p$-value of Fisher's test which adds $y_{0C}$ successes out of $n_{0C}$ external patients to the control arm. Moreover, under Jeffrey's independent prior specifications $\pi^{0,U}$, i.e., $a_C=b_C=a_T=b_T=0.5$, the Bayesian test approximates the frequentist unconditional test, as \citep{howa1998}
\begin{equation*}
P^{\pi^{0,U}}[\theta_T-\theta_C \leq 0| y_C, y_T] \approx \Phi \left[ (y_C(n_T-y_T)- y_T(n_C-y_C)) \sqrt{\frac{y_C +(n_C-y_C) + y_T +(n_T-y_T)}{n_C n_T (y_C+y_T) (n_C+n_T-y_C-y_T)}} \right],\\
\end{equation*}
and analogous considerations to the Fisher's test situation can be made for $P^{\pi^{U}}[\theta_T-\theta_C \leq 0| y_C, y_T]$ with $a_C=y_{0C}+0.5$, $b_C=n_{0C}-y_{0C} +0.5$, $a_T=0.5$, and $b_T=0.5$.
We focus in the following on the unconditional test, so $\pi^0=\pi^{0,U}$, and $\pi=\pi^{U}$, i.e., we adopt Jeffrey's independent priors. The debate on conditional versus unconditional testing is a long standing one. Here we focus on the unconditional test as it better approximates the nominal level under the null, and is induced by a more widely used prior in the Bayesian context.
Let
$$C(y_C, y_T) = P^{\pi}[\theta_T-\theta_C \leq 0| y_C, y_T]-P^{\pi^0}[\theta_T-\theta_C \leq 0| y_C, y_T].$$ 
Recall that the FD rejects $H_0$ when $P^{\pi^0}[\theta_T-\theta_C \leq 0| y_C, y_T] \leq \kappa$. Analogously, the BD rejects $H_0$ when $P^{\pi}[\theta_T-\theta_C \leq 0| y_C, y_T] \leq \gamma$. Therefore, the BD is obtained under $\pi^0$ by taking 
$$\kappa^{BD}(y_C,y_T)= \kappa - C(y_C, y_T).$$ 
Note that $\kappa-C(y_C, y_T)$ may be above 1 or below 0. This creates in practice no issues, as the condition $P^{\pi^0}[\theta_T-\theta_C \leq 0| y_C, y_T] \leq (\kappa - C(y_C, y_T))$ remains well defined and can be used to obtain a test decision.
Note that $C(y_C, y_T)$ can also be obtained analytically via the recursions \citep{howa1998}
\begin{align}
\label{eq:binomialord1} P^{\pi}[\theta_T-\theta_C \leq 0| y_C, y_T]&=P^{\pi^0}[\theta_T-\theta_C \leq 0| y_C, y_T] +\\
\nonumber &\sum^{y_{0C}-1}_{i=0} \frac{g(a_C+y_C+i,b_C+n_C-y_C,a_T+y_T,b_T+n_T-y_T)}{a_C+y_C+i} - \\
\nonumber &\sum^{n_{0C}-y_{0C}-1}_{i=0} \frac{g(a_C+y_C+y_{0C},b_C+n_C-y_C+i,a_T+y_T,b_T+n_T-y_T)}{b_C+n_C-y_C+i},
\end{align}
where $g(a,b,c,d)=(\Gamma(a+b)\Gamma(c+d)\Gamma(a+c)\Gamma(b+d))/(\Gamma(a)\Gamma(b)\Gamma(c)\Gamma(d)\Gamma(a+b+c+d))$.
From Equation (\ref{eq:binomialord1}), we can see that, all other quantities being equal, an increase in the number of external successes in the control arm $y_{0C}$ leads to a higher posterior probability of the null, making the test more conservative as it is less likely to satisfy the Bayesian condition for rejection $P^{\pi}[\theta_T-\theta_C \leq 0| y_C, y_T]\leq \gamma$, for any fixed $\gamma$. Analogously, for a fixed prior, an increase in the current number of successes also leads to a more conservative test as \citep{howa1998}
\begin{align*}
 P^{\pi}[\theta_T-\theta_C \leq 0| (y_C+1), y_T]=&  P^{\pi}[\theta_T-\theta_C \leq 0| y_C, y_T] +\\ & l(a_C+y_C+y_{0C},b_C+n_C-y_C+n_{0C}-y_{0C},a_T+y_T,b_T+n_T-y_T),
\end{align*}
where $l(c,d,a,b)=(\Gamma(a+b)\Gamma(c+d)\Gamma(a+c)\Gamma(b+d-1))/(\Gamma(a)\Gamma(b)\Gamma(c+1)\Gamma(d)\Gamma(a+b+c+d-1))$.
This suggests that an analogue of the CD for Normal outcomes can be proposed for binomial outcomes.  
The CDC is here obtained as
\begin{equation*}
  \kappa^{CDC}(y_C,y_T) = 
    \max(\min(\kappa^{BD}(y_C,y_T), \alpha^{UP}),\alpha^{LOW}), 
\end{equation*}

Let $\mu_C=y_{0C}/n_{0C}$, i.e., the prior proportion of successes in the control arm. Analogously to the Normal outcomes case, the CDD can be obtained by 
\begin{eqnarray*}
  \kappa^{CDD}(\bar{y}_C) &=& \max\left(\min\left(\left( \frac{\kappa}{\kappa^{BD}(y_C,y_T)}\right)^w \kappa^{BD}(y_C,y_T), \alpha^{UP}\right),\alpha^{LOW}\right) \\
\nonumber w &=&  \min\left( \left(\frac{|{y}_C /n_C-  \mu_C|}{t \sqrt{\hat{\rho}(1-\hat{\rho})\sqrt{1/n_C+1/n_{0C}} }}\right)^p,1 \right)  
\end{eqnarray*}
for $t>0$ and $p\geq0$, and where $\hat{\rho}=({y}_C+y_{0C})/(n_C+n_{0C})$ is the estimated common proportion of successes. Note, again, that we have implicitly adopted a normal approximation to estimate observed prior-data conflict. 

\section{Additional simulations}
\label{sec:addsim}

\subsection{Normal outcomes}
\label{sec:addsimnormal}

\begin{figure}[h!]
   \centering
   \includegraphics[width=1\textwidth]{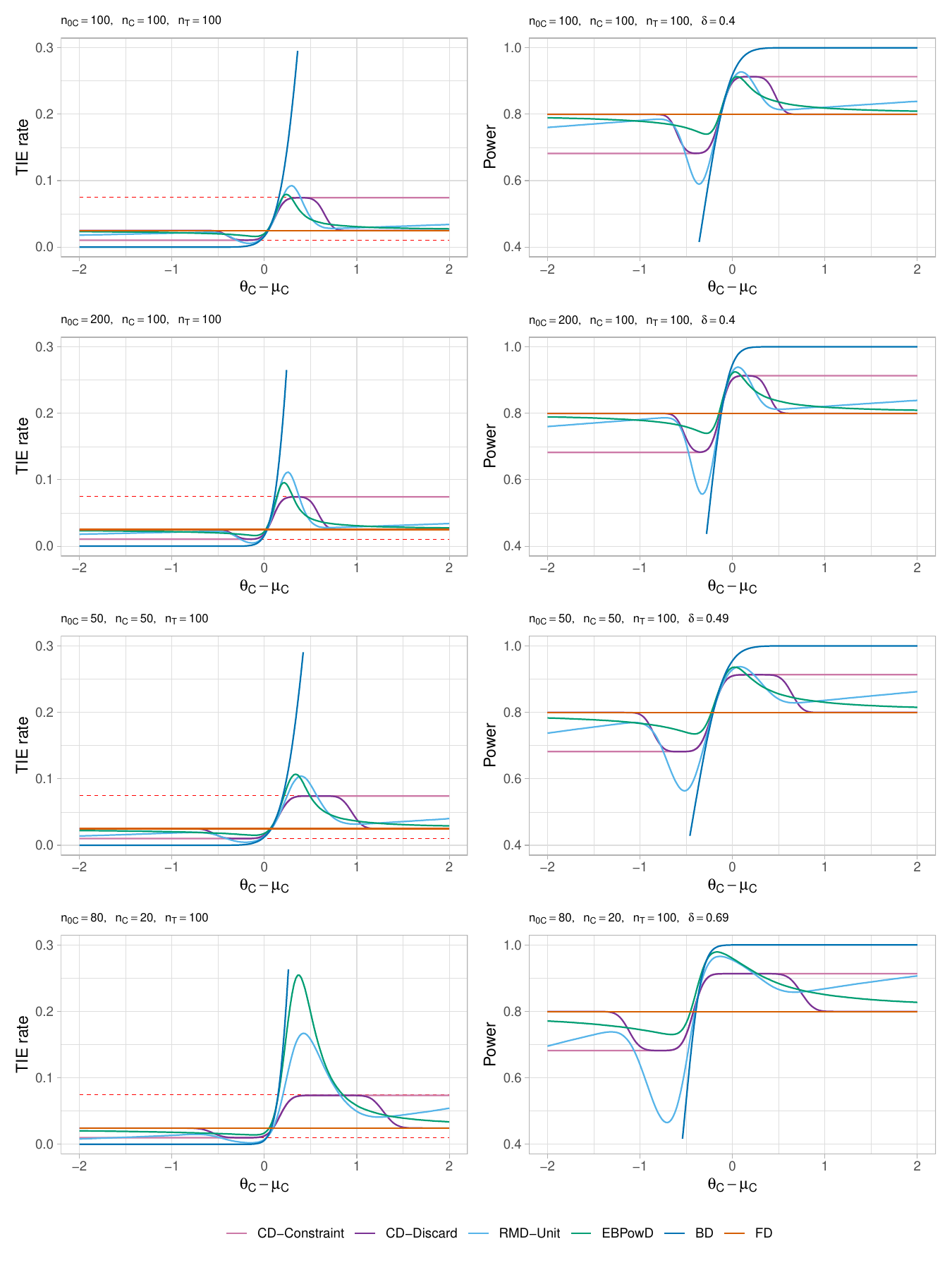}
   \caption{TIE rate and Power for different borrowing approaches as a function of the conflict $\theta_C-\mu_C$. Normal outcomes with known variance $\sigma^2=1$. Large current sample size scenarios: the current sample size and informative prior effective sample size are varied in each row. $t = 4$ and $p = 4$ for the CDD. $\alpha^{UP}=0.075$, $\alpha^{LOW}=0.01$.}
\label{fig:simnormlarge}
\end{figure}

\begin{figure}[h!]
   \centering
   \includegraphics[width=1\textwidth]{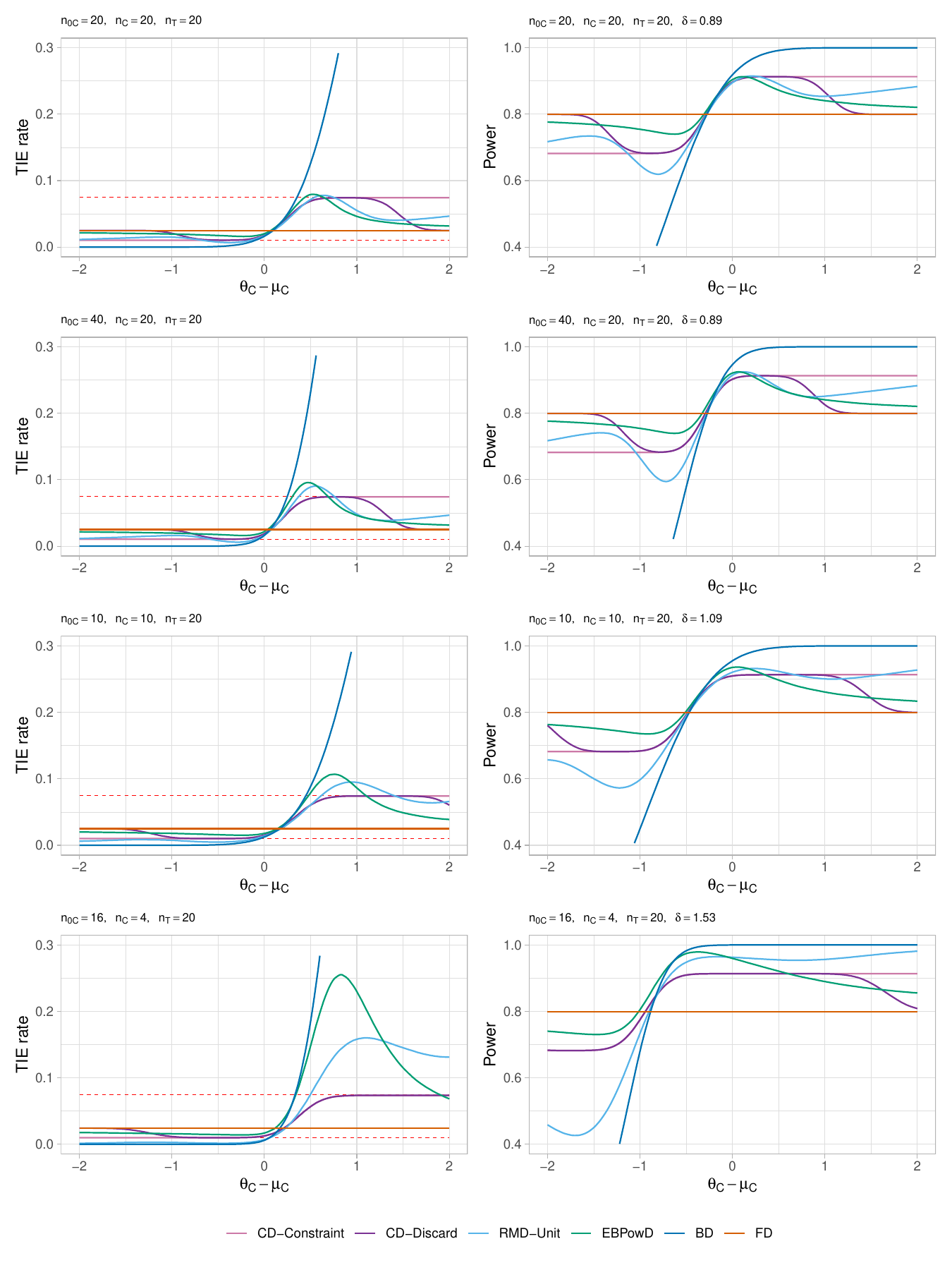}
   \caption{TIE rate and Power for different borrowing approaches as a function of the conflict $\theta_C-\mu_C$. Normal outcomes with known variance $\sigma^2=1$. Small current sample size scenarios: the current sample size and informative prior effective sample size are varied in each row. $t = 4$ and $p = 4$ for the CDD. $\alpha^{UP}=0.075$, $\alpha^{LOW}=0.01$.}
\label{fig:simnormsmall}
\end{figure}

\begin{figure}[h!]
   \centering
   \includegraphics[width=1\textwidth]{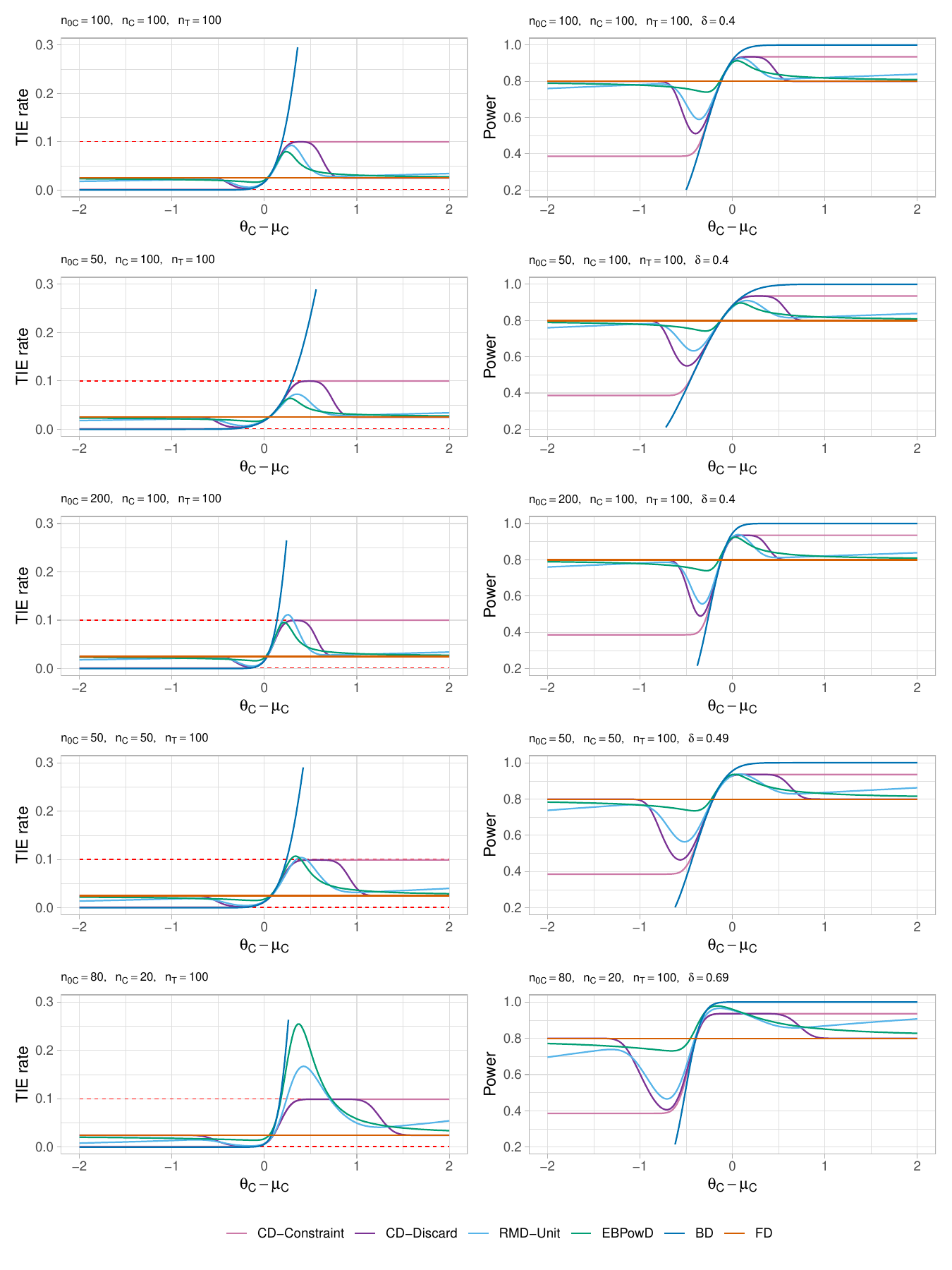}
   \caption{TIE rate and Power for different borrowing approaches as a function of the conflict $\theta_C-\mu_C$. Normal outcomes with known variance $\sigma^2=1$. Large current sample size scenarios: the current sample size and informative prior effective sample size are varied in each row. $t = 4$ and $p = 4$ for the CDD. $\alpha^{UP}=0.1$, $\alpha^{LOW}=0.001$.}
\label{fig:simnormlargewider}
\end{figure}

\begin{figure}[h!]
   \centering
   \includegraphics[width=1\textwidth]{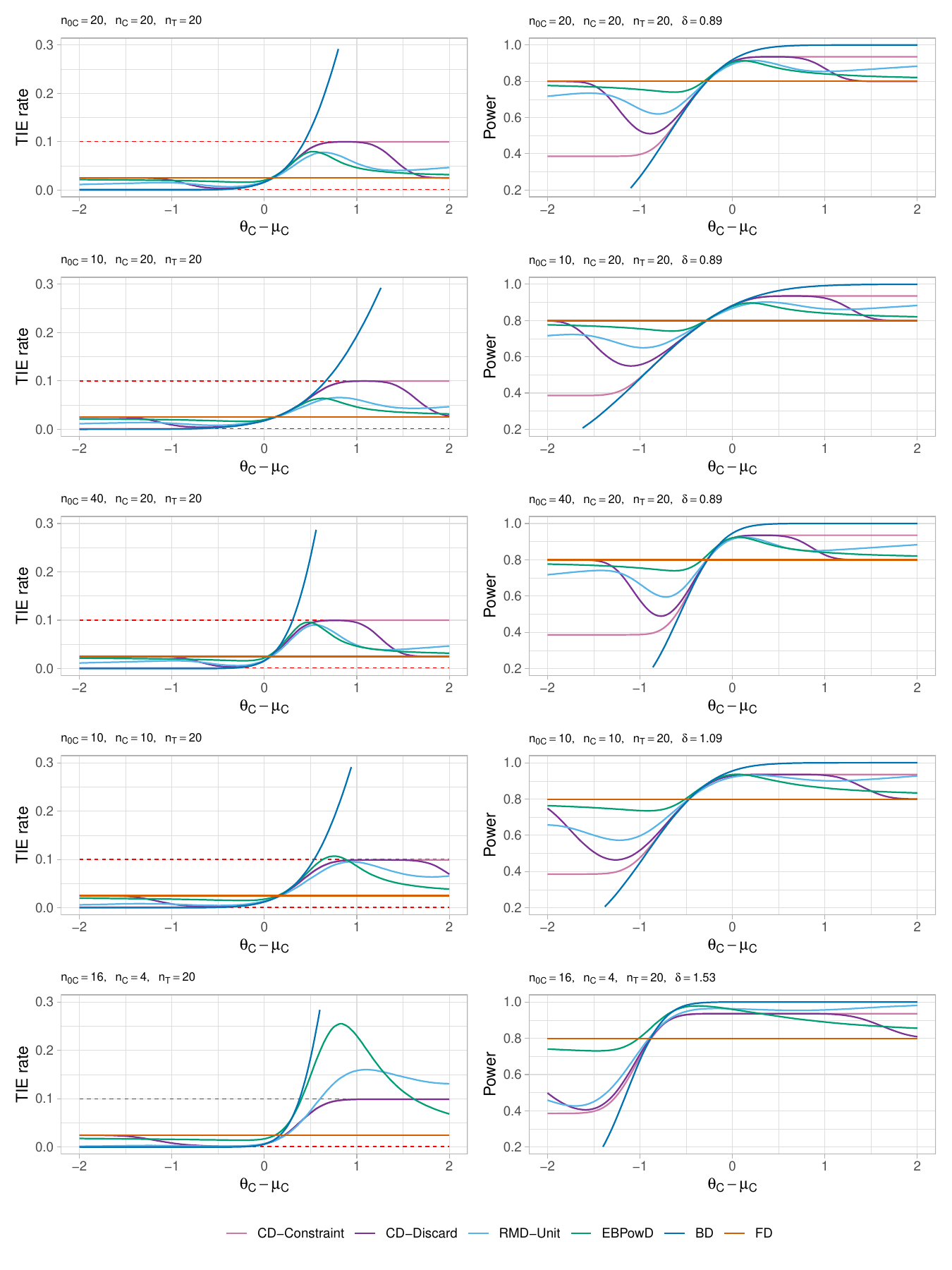}
   \caption{TIE rate and Power for different borrowing approaches as a function of the conflict $\theta_C-\mu_C$. Normal outcomes with known variance $\sigma^2=1$. Small current sample size scenarios: the current sample size and informative prior effective sample size are varied in each row. $t = 4$ and $p = 4$ for the CDD. $\alpha^{UP}=0.1$, $\alpha^{LOW}=0.001$.}
\label{fig:simnormsmallwider}
\end{figure}

\begin{figure}[h!]
   \centering
   \includegraphics[width=1\textwidth]{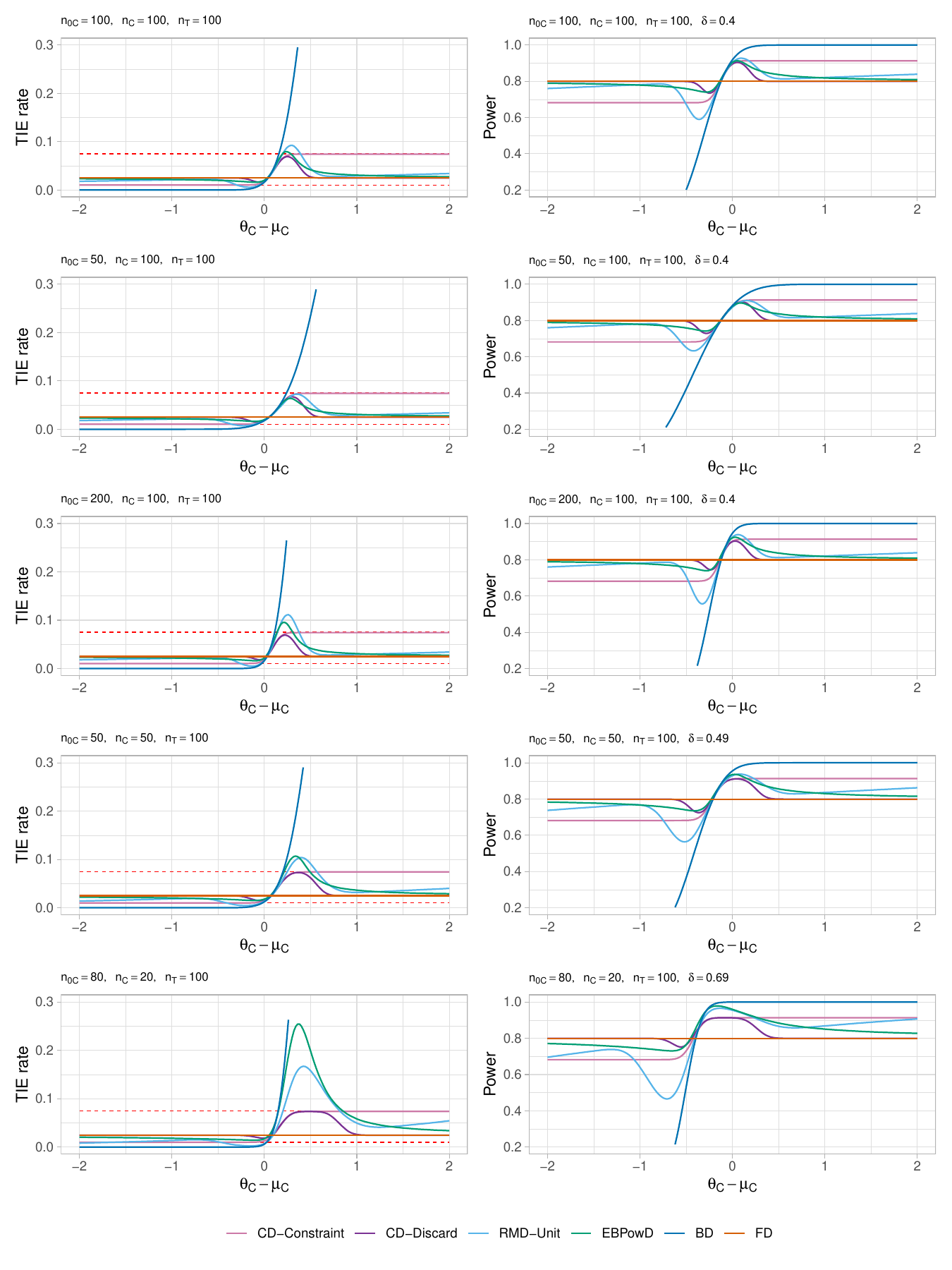}
   \caption{TIE rate and Power for different borrowing approaches as a function of the conflict $\theta_C-\mu_C$. Normal outcomes with known variance $\sigma^2=1$. Large current sample size scenarios: the current sample size and informative prior effective sample size are varied in each row. $t = 2$ and $p = 4$ for the CDD. $\alpha^{UP}=0.075$, $\alpha^{LOW}=0.01$.}
\label{fig:simnormlarget2}
\end{figure}

\begin{figure}[h!]
   \centering
   \includegraphics[width=1\textwidth]{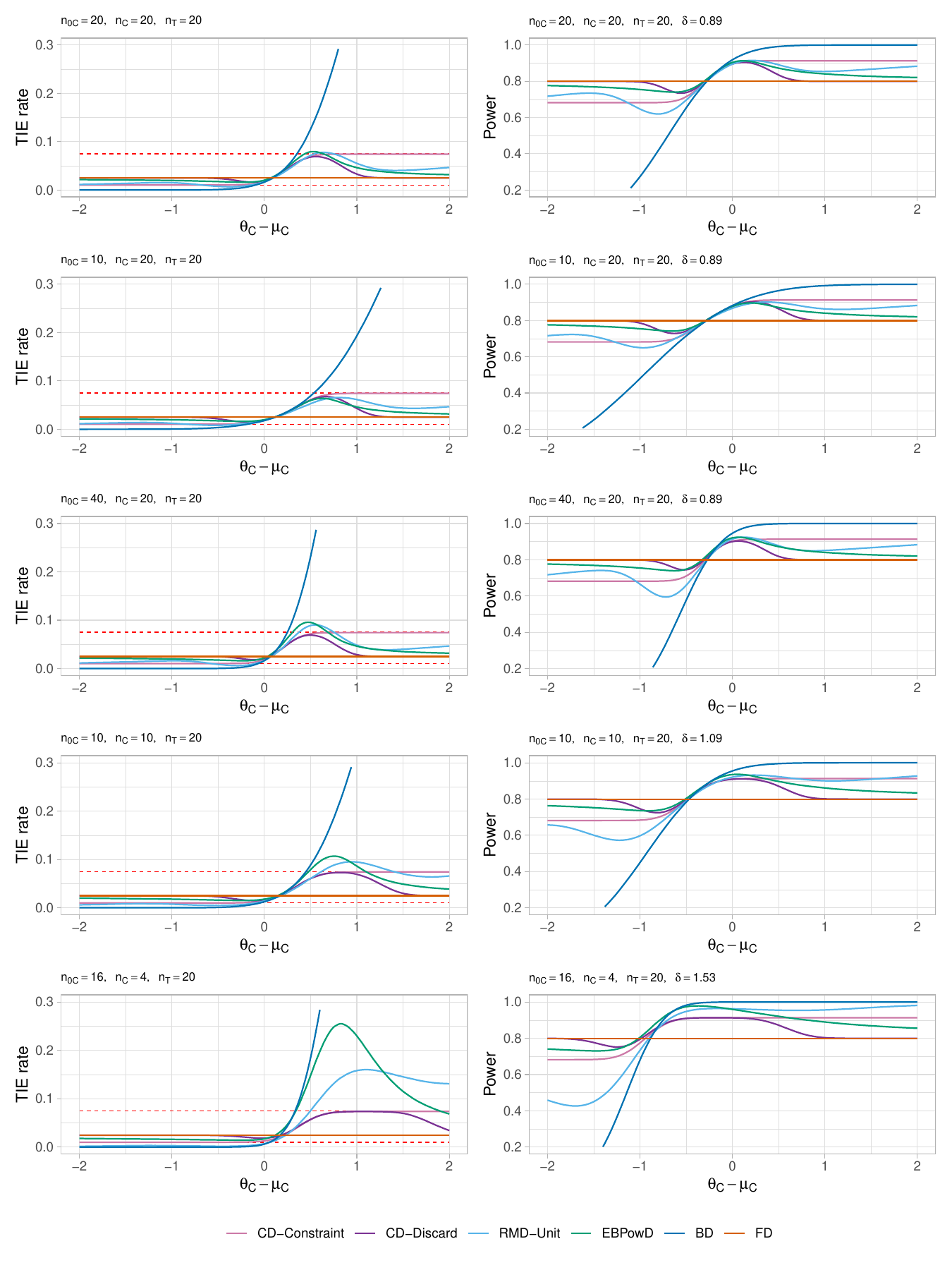}
   \caption{TIE rate and Power for different borrowing approaches as a function of the conflict $\theta_C-\mu_C$. Normal outcomes with known variance $\sigma^2=1$. Small current sample size scenarios: the current sample size and informative prior effective sample size are varied in each row. $t = 2$ and $p = 4$ for the CDD. $\alpha^{UP}=0.075$, $\alpha^{LOW}=0.01$.}
\label{fig:simnormsmallt2}
\end{figure}

Here we follow the same simulation set-up as the main text, but investigate situations in which the sample size of the control arm $n_C$ is equal, double, or smaller than that of the treated arm $n_T$, and, in the latter case, external control information is added to compensate for such imbalance, i.e., $n_{0C}=n_T-n_C$. Figures \ref{fig:simnormlarge} and \ref{fig:simnormsmall} show the results for various scenarios. We observe that the reduction in the number of current control samples leads in general to larger TIE rate inflation for the RMD-Unit and the EBPowD. For the CDs, on the other hand, having fixed the same cap on TIE rates across the different scenarios, this is no longer the case. The CDC has relatively similar behavior across the different scenarios, while the CDD accepts conflict on a wider range of current parameter values when the current sample size is smaller, consistently with the observation that there is more uncertainty about whether conflict is actually present. In our example the cap on TIE rate inflation also automatically caps the influence of 'dominating' prior information, i.e., when $n_{0C}$ is much larger than the current sample size, without prior discounting. Note also that the CDC switches between the BD and the constrained TIE rate rule more quickly when the current sample size is large.
In Figures \ref{fig:simnormlargewider} and \ref{fig:simnormsmallwider}, we additionally varied $\alpha^{UP}$ and $\alpha^{LOW}$, setting them to 0.001 and 0.1, respectively, meaning that we allow larger type I error rate inflation and much larger power loss. The CDs follow in this case the BD for a wider range of conflict values. $\alpha^{UP}$ and $\alpha^{LOW}$ could also be set closer the $\alpha=0.025$: as noted in the main text, such choice would reduce the power gains to the point that, if $\alpha^{UP}=0.025$, only power losses would be observed.
In Figures \ref{fig:simnormlarget2} and \ref{fig:simnormsmallt2}, we vary $t$, instead, by setting it equal to 2. Here, the CDD borrows for a much smaller range of parameter values, a behavior which is particularly pronounced in the small sample size scenarios. 

\paragraph{Calibration to borrowing}
\label{sec:calibration2borr}

\begin{figure}[h!]
   \centering
   \includegraphics[width=1\textwidth]{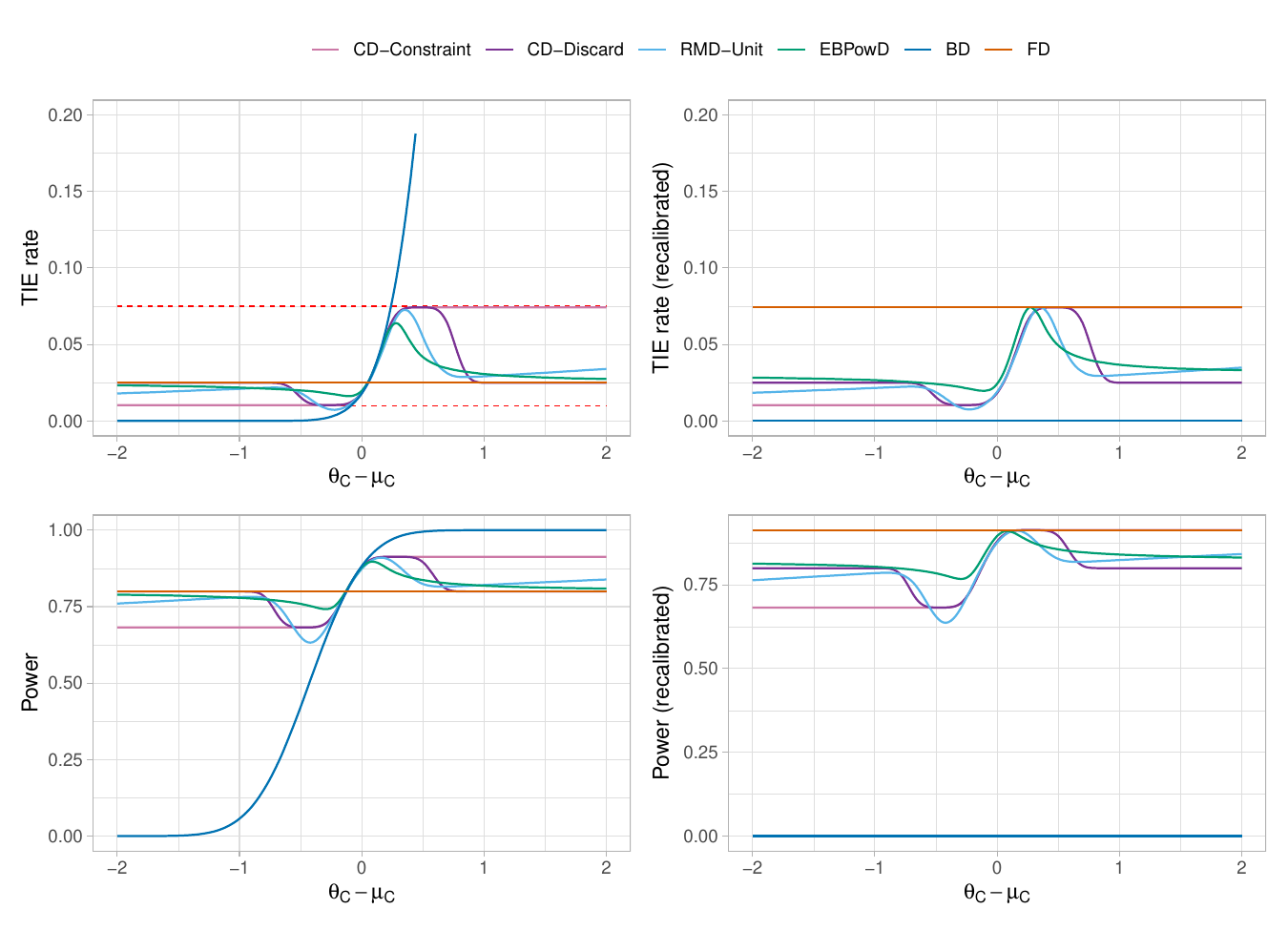}
   \caption{TIE rate and Power (left) and recalibrated TIE rate and Power (right) at $\delta=0.4$ for different borrowing approaches as a function of the conflict $\theta_C-\mu_C$. Normal outcomes with known variance $\sigma^2=1$, $n_T=n_C=100$. Informative prior $\pi_C=N(0,\sigma_C=\sigma/\sqrt{50})$. The dashed horizontal lines correspond to the TIE rate boundaries for the CD approaches, $\alpha^{LOW}=0.01$ and $\alpha^{UP}=0.075$. $t = 4$ and $p = 4$ for the CDD.}
\label{fig:sim2}
\end{figure}

To allow a fair comparison of power, following \cite{kopp2024} we recalibrate the TIE rate of all approaches to have the same maximum as the CDs within the range of conflict values considered. Note that, while in our case the conflict range has been chosen to allow visualization of the main borrowing dynamics, in practical applications its selection should be driven by considerations on a 'feasible' range for the control arm parameter.  We speculate that no dynamic approach can achieve at any point the same recalibrated power as the FD due to the fact that all dynamic borrowing approaches deviate from the UMP test at any given $\theta_C$ as they adapt to the observed conflict. Without adaptation, on the other hand, no calibration is possible as the maximum TIE rate would be 1. The CDs reach however power differences closer to zero, and for a wider range of conflict values, compared to the EBPowD and RMD-Unit, as can be observed in Table \ref{tab:recalibration}, Figure \ref{fig:sim1} and Supplementary Figure \ref{fig:sim2}. This is due to the fact that the CDs have stable TIE rate and therefore power over a wider range of $\theta_C$ parameter values. Note that for the BD, as noted, TIE rate can only be controlled below 0.075 if the null hypothesis is never rejected: hence, the difference in TIE rate is approximately -0.075 (being 0.075 the TIE rate of the calibrated FD), and the power difference analogously follows from the power of the FD.

\subsection{Binomial outcomes}
\label{sec:addsimbinomial}

\begin{figure}[h!]
   \centering
   \includegraphics[width=1\textwidth]{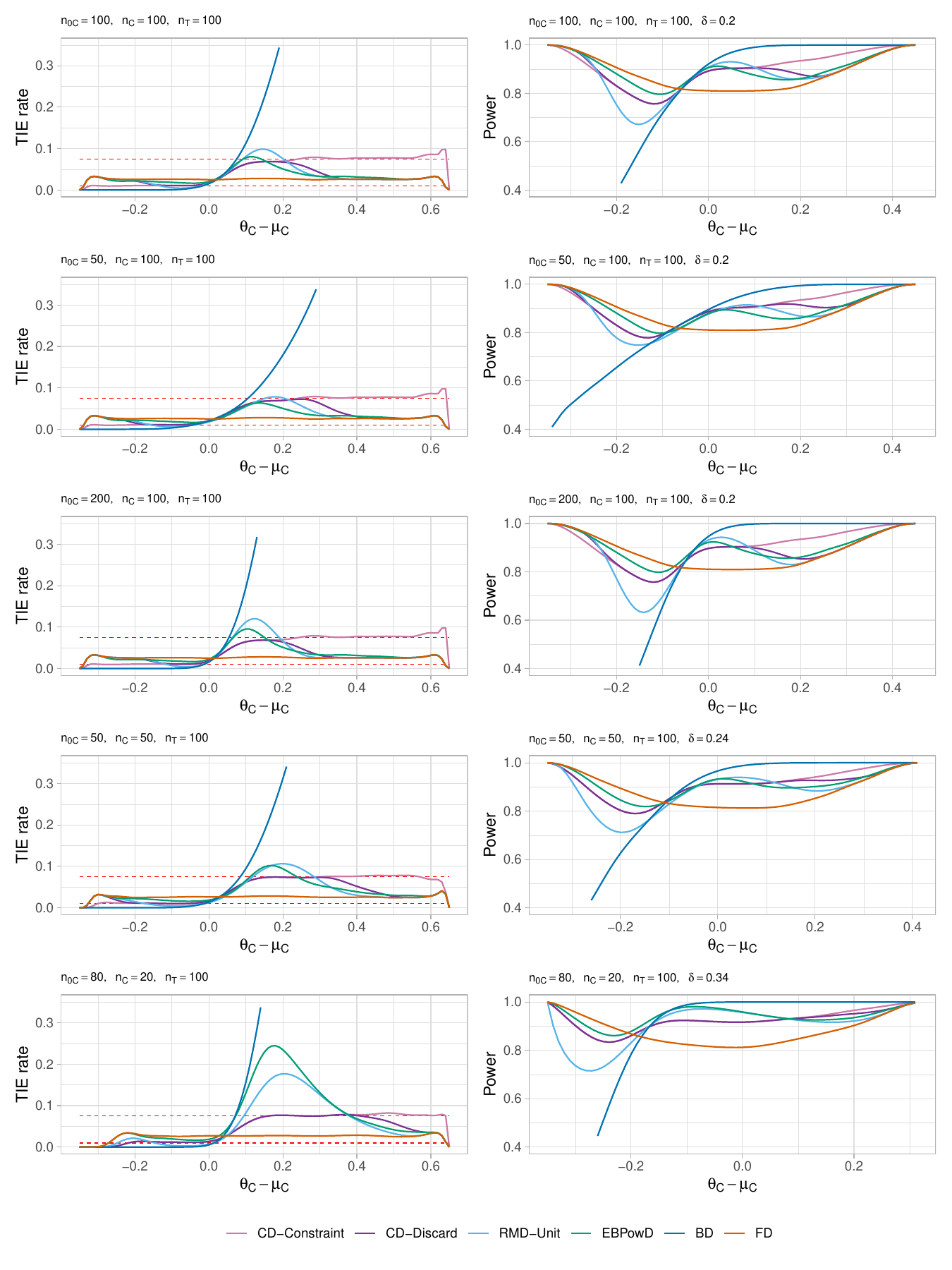}
   \caption{TIE rate and Power for different borrowing approaches as a function of the conflict $\theta_C-\mu_C$. Binomial outcomes. Large current sample size scenarios: the current sample size and informative prior effective sample size are varies in each row. $t = 4$ and $p = 4$ for the CDD. $\alpha^{UP}=0.075$, $\alpha^{LOW}=0.01$.}
\label{fig:simbinomlarge}
\end{figure}

\begin{figure}[h!]
   \centering
   \includegraphics[width=1\textwidth]{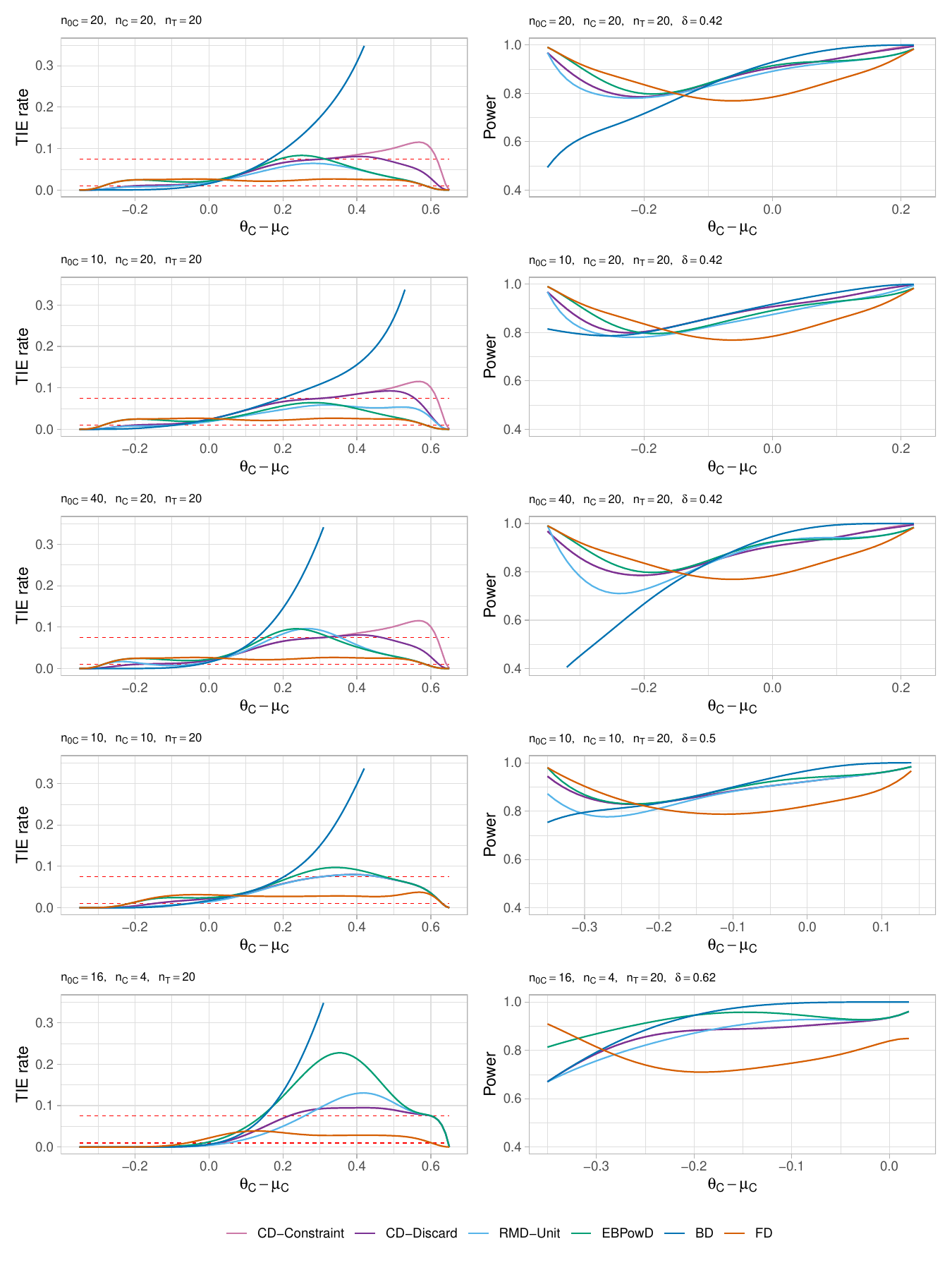}
   \caption{TIE rate and Power for different borrowing approaches as a function of the conflict $\theta_C-\mu_C$. Binomial outcomes. Small current sample size scenarios: the current sample size and informative prior effective sample size are varied in each row. $t = 4$ and $p = 4$ for the CDD. $\alpha^{UP}=0.075$, $\alpha^{LOW}=0.01$.}
\label{fig:simbinosmmall}
\end{figure}

\begin{figure}[h!]
   \centering
   \includegraphics[width=1\textwidth]{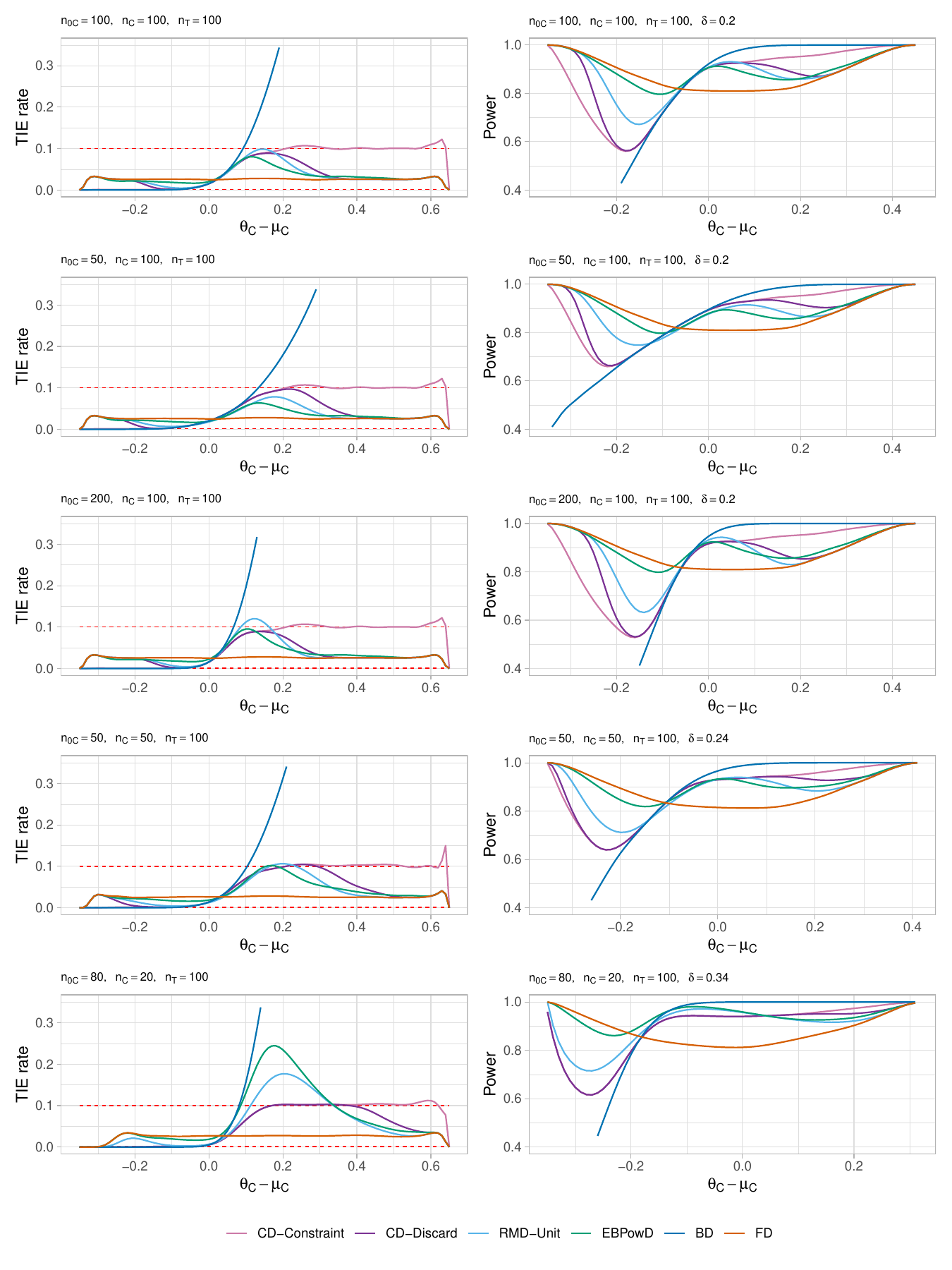}
   \caption{TIE rate and Power for different borrowing approaches as a function of the conflict $\theta_C-\mu_C$. Binomial outcomes. Large current sample size scenarios: the current sample size and informative prior effective sample size are varied in each row. $t = 4$ and $p = 4$ for the CDD. $\alpha^{UP}=0.1$, $\alpha^{LOW}=0.001$.}
\label{fig:simbinomlargewider}
\end{figure}

\begin{figure}[h!]
   \centering
   \includegraphics[width=1\textwidth]{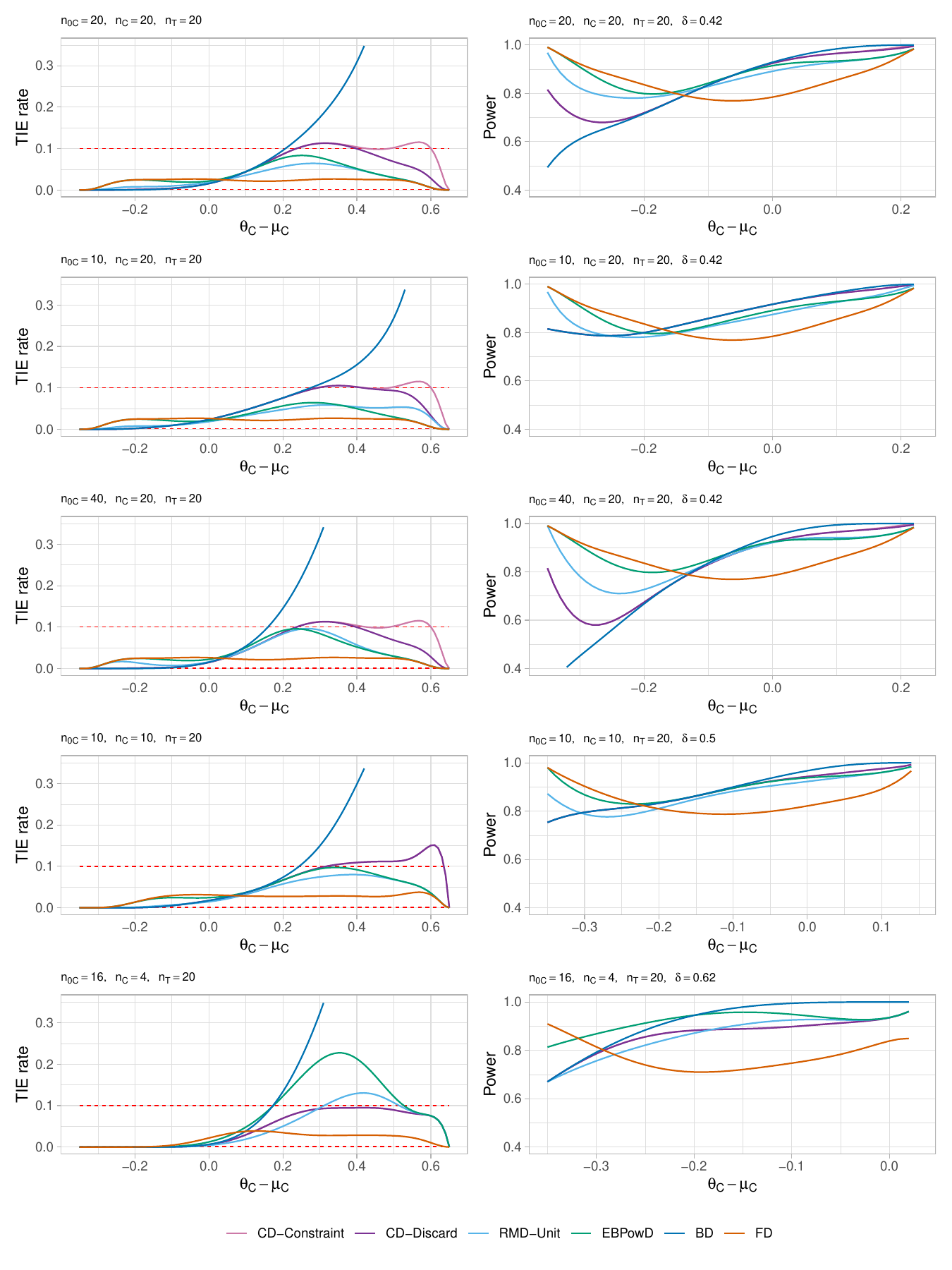}
   \caption{TIE rate and Power for different borrowing approaches as a function of the conflict $\theta_C-\mu_C$. Binomial outcomes. Small current sample size scenarios: the current sample size and informative prior effective sample size are varied in each row. $t = 4$ and $p = 4$ for the CDD. $\alpha^{UP}=0.1$, $\alpha^{LOW}=0.001$.}
\label{fig:simbinomsmallwider}
\end{figure}

\begin{figure}[h!]
   \centering
   \includegraphics[width=1\textwidth]{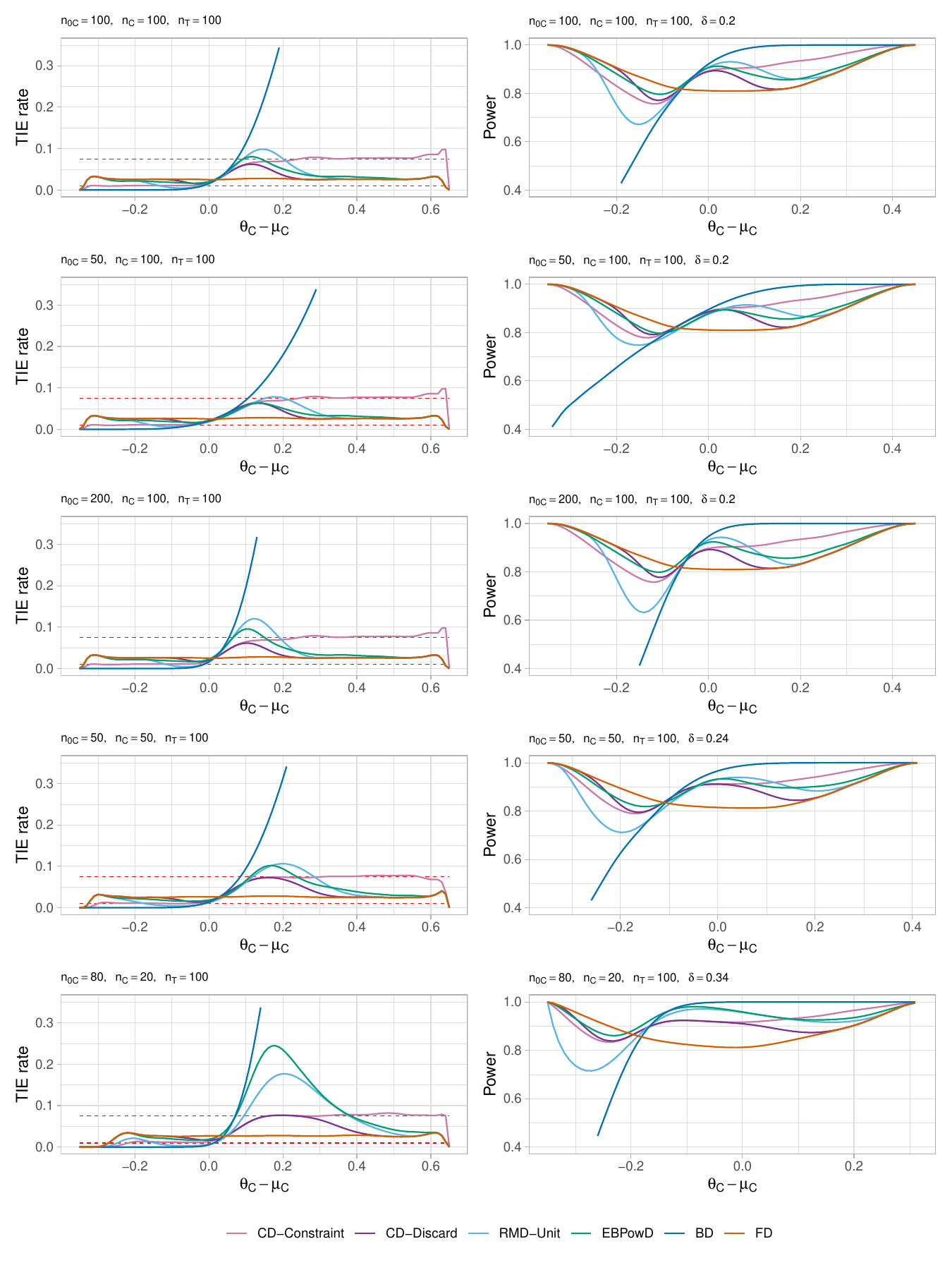}
   \caption{TIE rate and Power for different borrowing approaches as a function of the conflict $\theta_C-\mu_C$. Binomial outcomes. Large current sample size scenarios: the current sample size and informative prior effective sample size are varies in each row. $t = 2$ and $p = 4$ for the CDD. $\alpha^{UP}=0.075$, $\alpha^{LOW}=0.01$.}
\label{fig:simbinomlarget2}
\end{figure}

\begin{figure}[h!]
   \centering
   \includegraphics[width=1\textwidth]{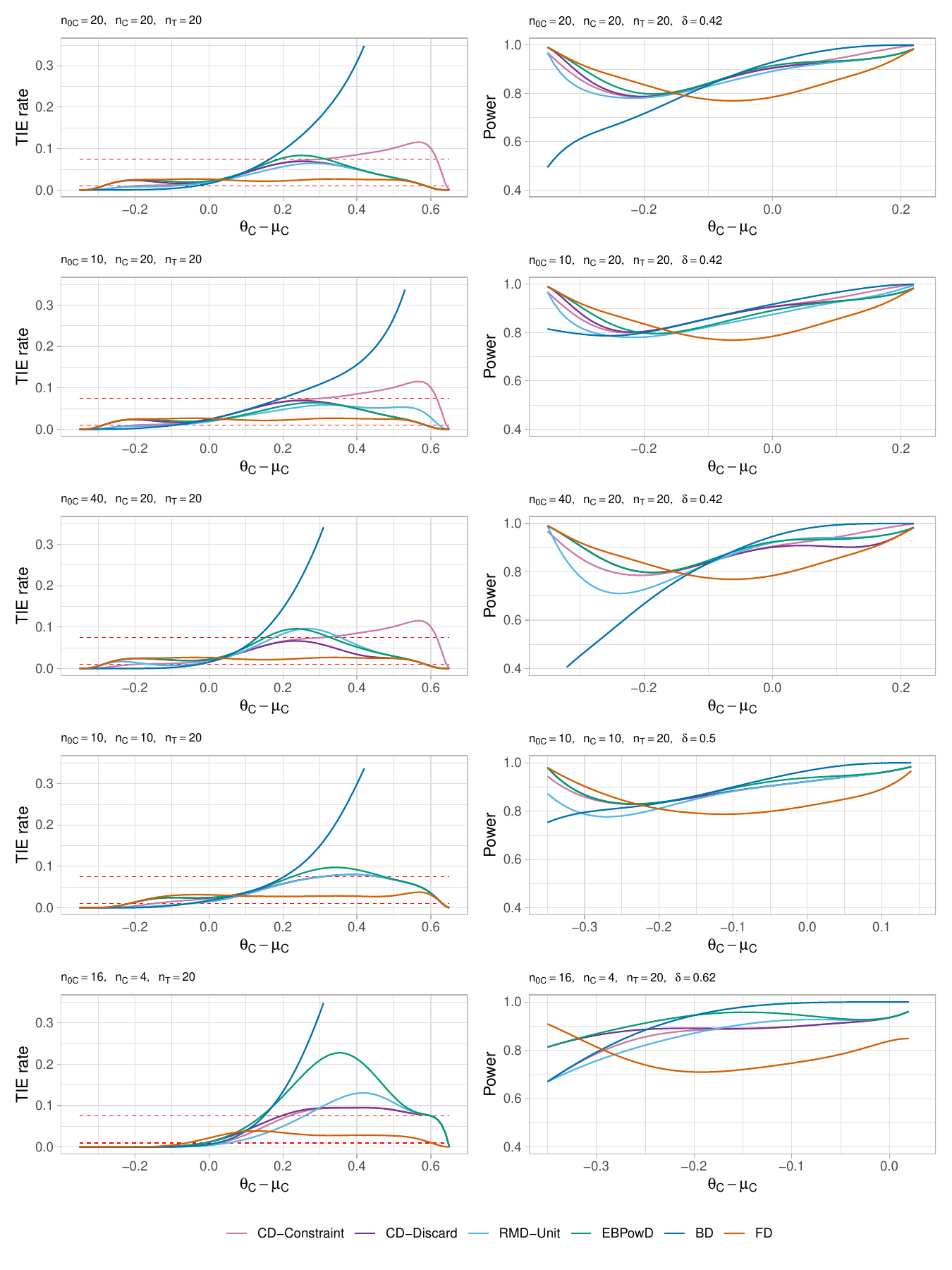}
   \caption{TIE rate and Power for different borrowing approaches as a function of the conflict $\theta_C-\mu_C$. Binomial outcomes. Small current sample size scenarios: the current sample size and informative prior effective sample size are varied in each row. $t = 2$ and $p = 4$ for the CDD. $\alpha^{UP}=0.075$, $\alpha^{LOW}=0.01$.}
\label{fig:simbinomsmallt2}
\end{figure}

The simulation study is complemented with results for Binomial outcomes. We vary the sample sizes of each arm as for the Normal outcomes. The prior mean for the control arm is taken to be $\mu_C=0.35$, based on a umber of historical samples that once again mirrors the simulation setup for the Normal outcomes, i.e., half, double, the same, or complementing $n_C$. It is assumed that no information is available for the treatment arm, therefore we take $a_T=0.5$ and $b_T=0.5$.
The conflict $c$ is in this case varied between -0.35 and 0.65, therefore covering a range for $\theta_C$ between 0 and 1, and power is evaluated at a value $\delta$ which gives approximately 80\% power under the frequentist (no borrowing) approach. Again, we take $\alpha^{UP}=0.075$, $\alpha^{LOW}=0.01$, or $\alpha^{UP}=0.1$, $\alpha^{LOW}=0.001$, and $p=4$, and $t=4$ or $t=2$ for the CDD. Error rates are calculated via complete enumeration. Supplementary Figures \ref{fig:simbinomlarge}-\ref{fig:simbinomsmallt2} show the results.
Due to the discreteness of the Binomial and the use of the unconditional test, control of TIE rate of the CDD and CDC is only approximate, especially for small sample sizes. If this is a strict requirement the variation of the approach which uses Fisher's exact test is recommended. For larger sample sizes, the cap is better controlled. We observe patterns similar to those observed for normal outcomes: the CDs automatically cap the impact of prior information on test decisions under any combination of current and external sample size scenario, adaptation to the prespecified rules is faster when the current sample size is large, and wider ranges for TIE rate inflation and deflation allow to follow the BD decision for a larger range of conflict values.

\section{Binomial outcomes case study}
\label{sec:casestudy}

\begin{figure}
   \centering
   \includegraphics[width=1\textwidth]{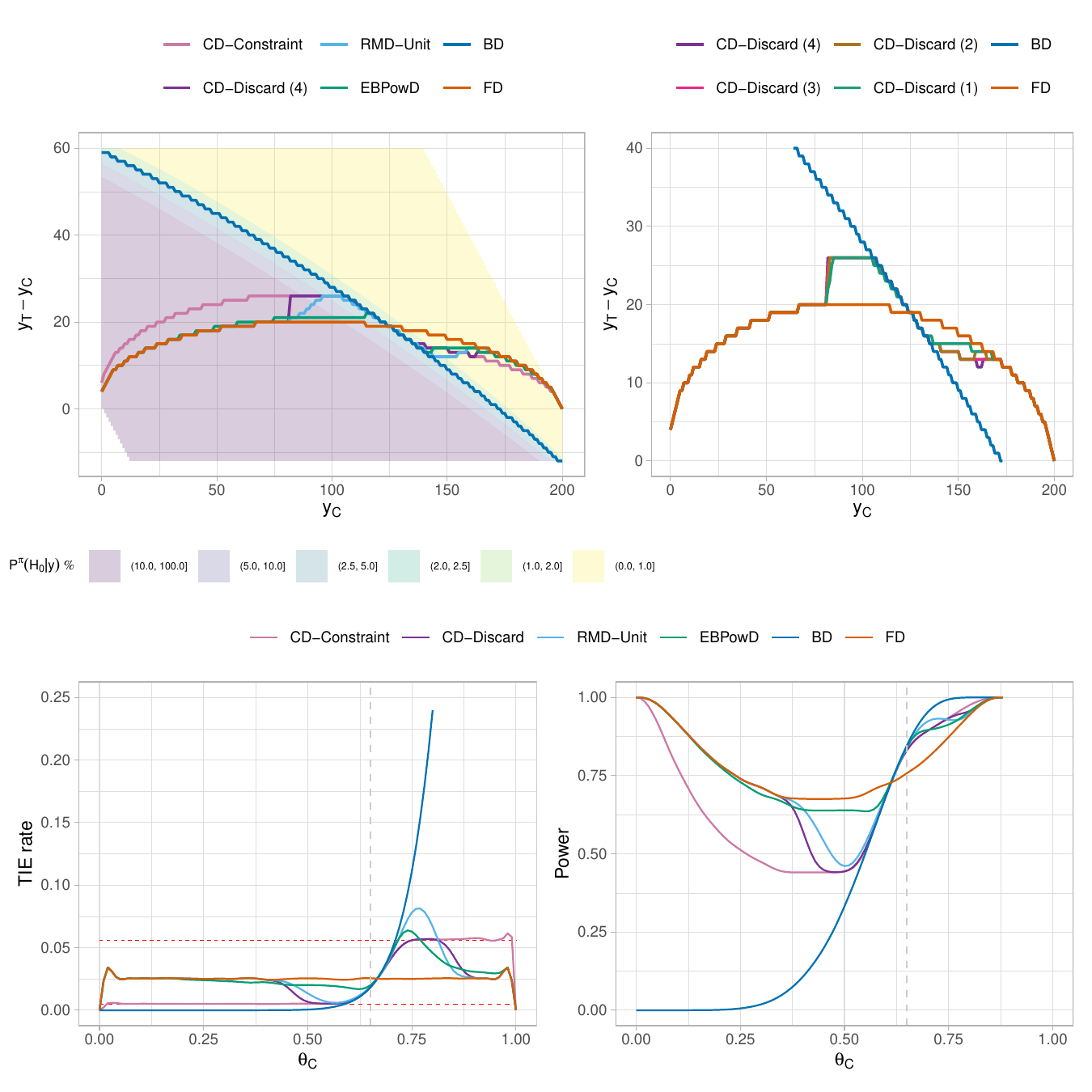}
   \caption{TIE rate and Power at $\delta=0.12$ for different borrowing approaches as a function of the conflict $\theta_C-\mu_C$. Binomial outcomes, $n_T=n_C=200$. Informative prior $\pi_C=Beta(65,35)$. The dashed horizontal lines correspond to the TIE rate boundaries for the CD approaches, $\alpha^{LOW}=0.005$ and $\alpha^{UP}=0.056$, $p=4$ and $t=4$ for the CDD. Simulation setup as in \cite{viele2014}.}
\label{fig:casestudyBinom}
\end{figure}

To illustrate the behavior of the method for binomial outcomes, we focus on the example in \cite{viele2014}. In particular, we take $n_C=n_T=200$ current subjects in each arm. The prior mean for the control arm is taken to be $\mu_C=0.65$, based on 100 historical subjects, i.e. $a_C=65$, $b_C=35$. It is assumed that no information is available for the treatment arm, therefore we take $a_T=0.5$ and $b_T=0.5$.
The conflict $c$ is in this case varied between -0.65 and 0.35, therefore covering a range for $\theta_C$ between 0 and 1, and power is evaluated at $\delta=0.12$. We take $\alpha^{UP}=0.056$, $\alpha^{LOW}=0.005$, corresponding to the TIE rates observed at approximately the 10\% and 90\% prior quantiles under the BD, and $p=4$ and $t=4$ for the CDD. Error rates are calculated via complete enumeration.
Results are shown in Figure \ref{fig:casestudyBinom}. For the CDC, as long as we do not risk increasing or decreasing the TIE rate above or below the pre-specified bounds, we reject according to the BD; for the CDD approach, such condition is complemented with a full discard when prior information is assessed to be `unacceptably surprising' compared to the current data, and such discard is achieved according to a speed determined by the parameter $p$. Focusing first on the conditional assessments, we can observe the impact that each approach has on the rejection region. For the chosen $\alpha^{UP}$ and $\alpha^{LOW}$, the CDs tend to favor more rejections for positive conflict and less for negative conflict, compared to the RMD-Unit and EBPowD. The borrowing parameter $p$ seems to not influence significantly the rejection region, providing relatively similar test decision thresholds when it varies between 2 and 4, while $p=1$ has a more evident impact mainly for positive conflict.
With respect to long-run properties, the CDs show a larger TIE rate inflation and power loss compared to the RMD-Unit and EBPowD. It would of course be possible to further constrain TIE rate if such inflation and/or deflation is considered unacceptable.  

\clearpage
\section{Supplementary Figures and Tables}

\begin{figure}[h!]
   \centering
   \includegraphics[width=1\textwidth]{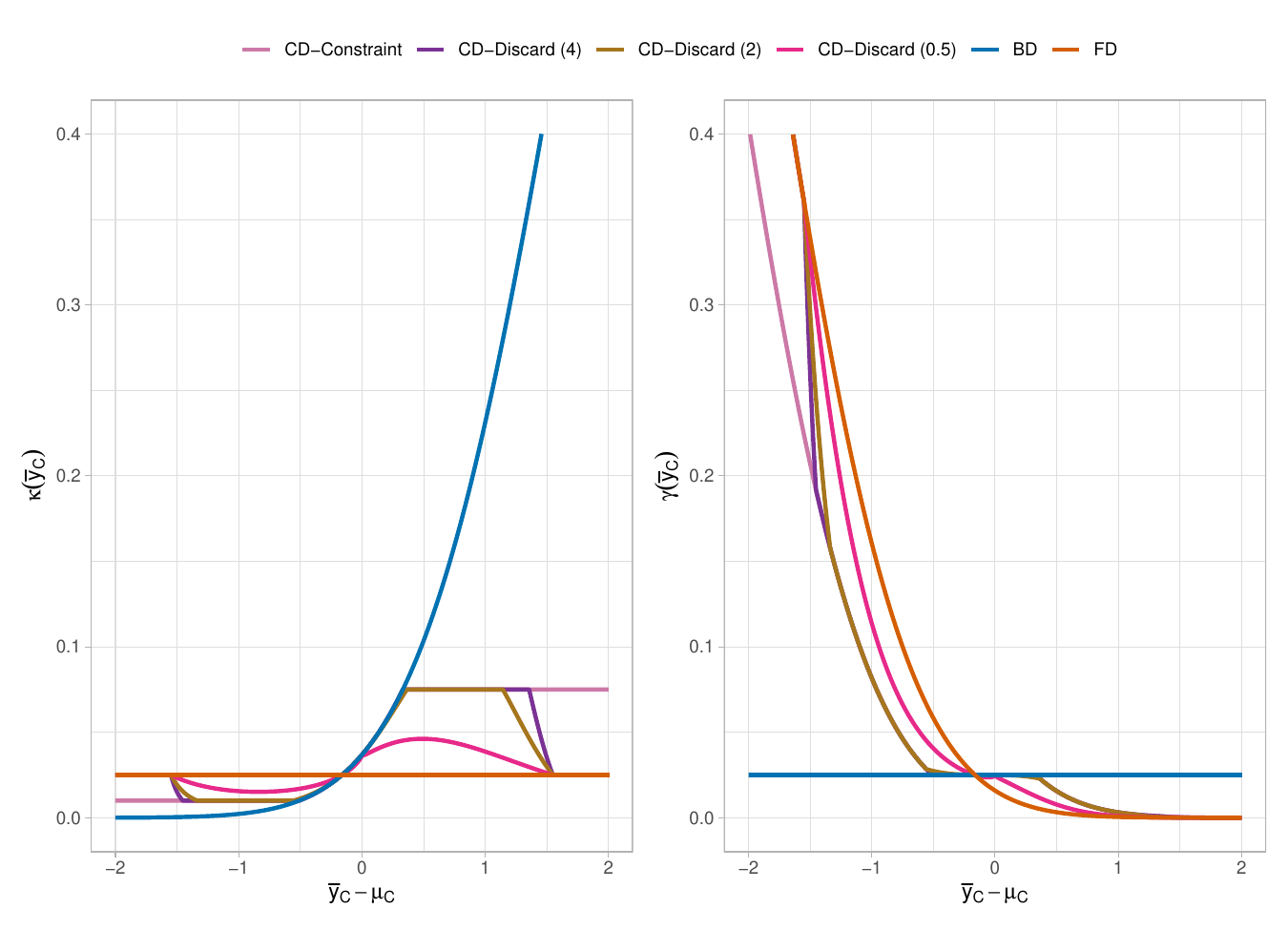}
   \caption{Frequentist test decision threshold (left), and Bayes test decision threshold (right) for different borrowing approaches as a function of the observed conflict $\bar{y}_C-\mu_C$. Normal outcomes with known variance $\sigma^2=1$, $n_T=n_C=20$. Informative prior $\pi_C=N(0,\sigma_C=\sigma/\sqrt{10})$, $t = 4$, and $p = {0.5,2,4}$ for the CDD approach.}
\label{fig:kappaCD}
\end{figure}

\begin{figure}[h!]
   \centering
   \includegraphics[width=1\textwidth]{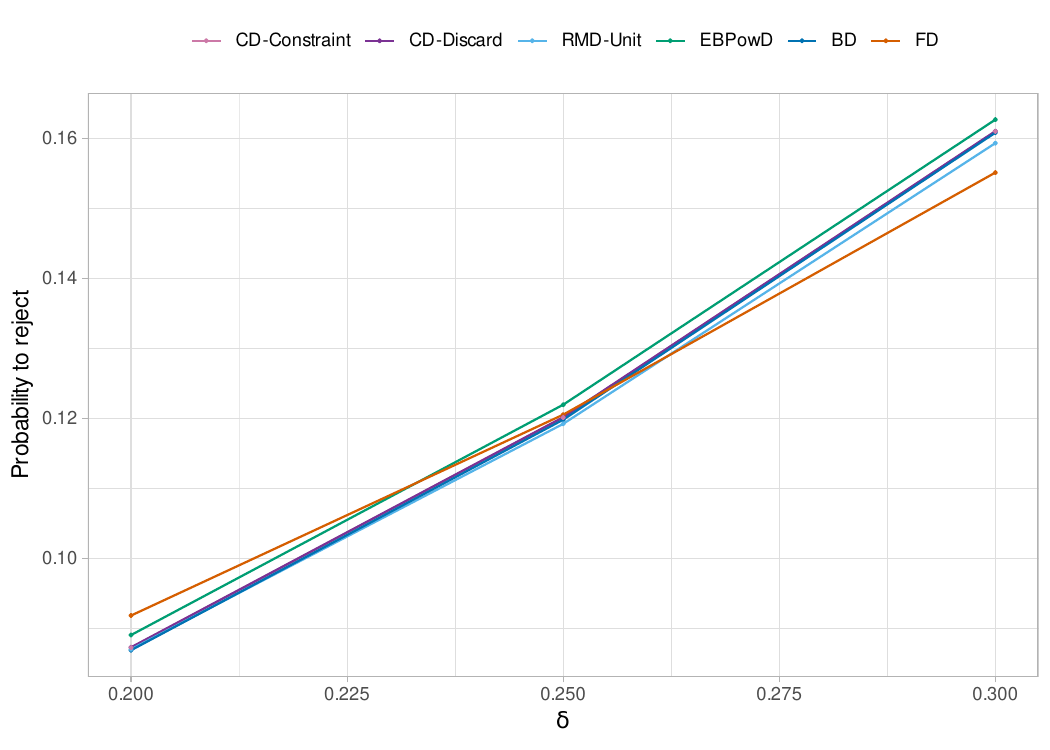}
   \caption{Power functions for $\theta_C=\mu_C$ for different borrowing approaches. Normal outcomes with known variance $\sigma^2=1$, $n_T=n_C=20$. Informative prior $\pi_C=N(0,\sigma_C=\sigma/\sqrt{10})$. For the CDs, $\alpha^{LOW}=0.01$ and $\alpha^{UP}=0.075$, and $t = 4$ and $p = 4$ for the CDD. The range of $\delta$ is chosen to show instances in which power of the robust borrowing approaches is lower or higher than both the FD and BD.}
\label{fig:EBpower}
\end{figure}

\begin{figure}[h!]
   \centering
   \includegraphics[width=1\textwidth]{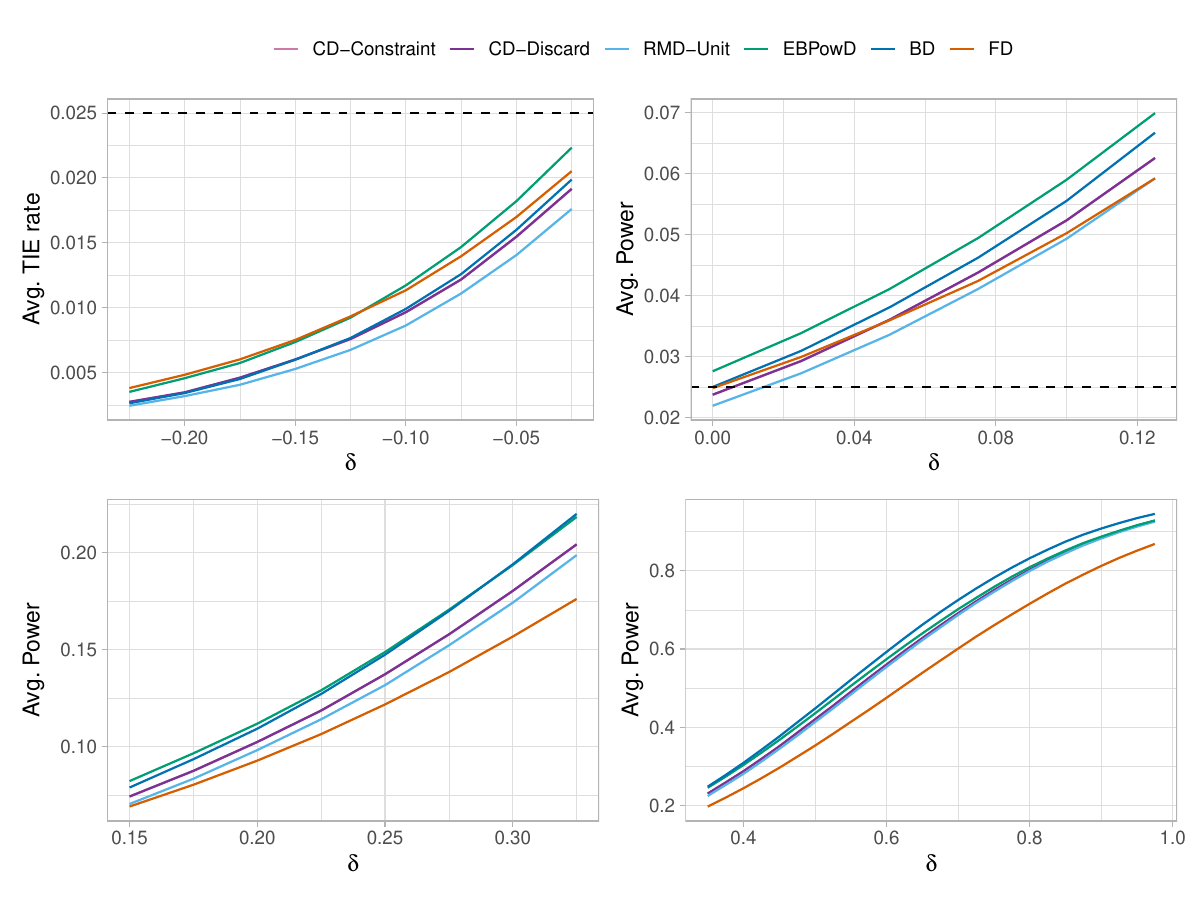}
   \caption{Average power function for the different borrowing approaches when adopting the informative prior $\pi_C=N(0,\sigma_C=\sigma/\sqrt{10})$ as sampling prior. Sampling and analysis prior coincide for the BD. Normal outcomes, known variance $\sigma^2=1$, $n_T=n_C=20$. For the CDs $\alpha^{LOW}=0.01$ and $\alpha^{UP}=0.075$, $t = 4$ and $p = 4$ for the CDD.}
\label{fig:avgOC}
\end{figure}

\begin{table}[ht]
\centering
\begin{tabular}{r|r|rrrrr|rrrrr|}
  \toprule
  \multicolumn{2}{c}{} & \multicolumn{5}{|c|}{$n_C=n_T=20$}  & \multicolumn{5}{c|}{$n_C=n_T=100$} \\      \cmidrule{3-12}
\multicolumn{2}{c}{}  &  \multicolumn{10}{|c|}{$\delta$}  \\ 
\multicolumn{2}{c|}{} & 0 & 0.25 & 0.5 & 0.75 & 1  & 0 & 0.1 & 0.2 & 0.3 & 0.4 \\ \midrule
& BD & 2.49 & 16.09 & 49.23 & 82.77 & 97.19 & 2.49 & 13.70 & 41.05 & 73.88 & 93.38 \\ 
&  FD & 2.48 & 12.17 & 35.32 & 66.03 & 88.44 & 2.48 & 10.57 & 29.34 & 56.52 & 80.74 \\ 
&  RMD-Unit & 2.11 & 13.85 & 44.24 & 77.94 & 95.06 & 2.20 & 12.27 & 37.82 & 70.13 & 90.99 \\ 
 \multirow{2}{*}{\parbox{3cm}{\centering Uncalibrated}} &  CDD & 2.28 & 14.06 & 43.75 & 77.06 & 94.67 & 2.28 & 11.99 & 36.16 & 67.60 & 89.40 \\ 
 & CDC   & 2.28 & 14.06 & 43.74 & 77.06 & 94.67 & 2.28 & 11.98 & 36.15 & 67.60 & 89.40 \\       \cmidrule{2-12}
&  EBPowD & 2.85 & 16.17 & 47.32 & 79.51 & 95.21 & 2.85 & 13.92 & 39.69 & 70.61 & 90.66 \\ 
    \multirow{2}{*}{\parbox{3cm}{\centering Calibrated to EBPowD}} &  CDD & 2.79 & 15.83 & 46.73 & 79.44 & 95.58 & 2.79 & 13.59 & 39.10 & 70.37 & 90.94 \\ 
  & CDC& 2.79 & 15.83 & 46.73 & 79.44 & 95.58 & 2.79 & 13.59 & 39.10 & 70.37 & 90.94 \\    \cmidrule{2-12}
    &  RMD-Unit (w=0.75)& 2.14 & 14.09 & 44.85 & 78.60 & 95.36 & 2.24 & 12.47 & 38.26 & 70.66 & 91.35 \\  
 \multirow{2}{*}{\parbox{3cm}{\centering Calibrated to RMD-Unit (w=0.75)}} &  CDD   & 2.23 & 14.36 & 45.15 & 78.80 & 95.53 & 2.31 & 12.67 & 38.43 & 70.81 & 91.54 \\
 & CDC & 2.22 & 14.34 & 45.12 & 78.78 & 95.53 & 2.30 & 12.64 & 38.39 & 70.79 & 91.54 \\        \cmidrule{2-12}
    & RMD-Unit (w=0.5)& 2.03 & 13.05 & 41.99 & 75.46 & 93.87 & 2.10 & 11.64 & 36.08 & 67.81 & 89.52 \\ 
 \multirow{2}{*}{\parbox{3cm}{\centering Calibrated to RMD-Unit (w=0.5)}}   &  CDD  & 2.13 & 13.31 & 42.26 & 75.63 & 94.03 & 2.19 & 11.86 & 36.27 & 68.02 & 89.79 \\ 
 & CDC  & 2.13 & 13.30 & 42.25 & 75.62 & 94.03 & 2.18 & 11.85 & 36.26 & 68.02 & 89.79 \\       \cmidrule{2-12}
    & RMD-Unit (w=0.25) & 2.04 & 12.36 & 39.07 & 71.95 & 92.11 & 2.05 & 10.86 & 33.44 & 64.11 & 86.91 \\  
  \multirow{2}{*}{\parbox{3cm}{\centering Calibrated to RMD-Unit (w=0.25)}}  & CDD  & 2.01 & 12.38 & 39.43 & 72.44 & 92.44 & 2.13 & 11.08 & 33.69 & 64.32 & 87.19 \\ 
 & CDC  & 2.01 & 12.38 & 39.42 & 72.44 & 92.44 & 2.13 & 11.08 & 33.69 & 64.31 & 87.19 \\ 
       \bottomrule
\end{tabular}
\caption{Average TIE rates (\%) and power (\%) for varying treatment effects $\delta$.  Normal outcomes with known variance $\sigma^2=1$, $n_C=n_T=20$ and $n_C=n_T=100$.  Averages are obtained by integrating over the informative prior for the control arm mean $\pi_C=N(0,\sigma_C=\sigma/\sqrt{n_{0C}})$, $n_{0C}=2n_C$, corresponding to the analysis prior for the BD. $t = 4$ and $p = 4$ for the CDD. In the uncalibrated scenario, for the CDs $\alpha^{LOW}=0.01$  and $\alpha^{UP}=0.075$, and the weight of the mixture prior $w=0.7$. In each calibrated scenario, $\alpha^{LOW}$ and $\alpha^{UP}$ for the CDC and CDD are set to the minimum and maximum TIE rate observed over the conflict range [-2,2] for the competitor approach.}
\label{tab:avgOC2}
\end{table}

\begin{table}[ht]
\centering
\begin{tabular}{r|r|rrrrr|rrrrr|}
  \toprule
  \multicolumn{2}{c}{} & \multicolumn{5}{|c|}{$n_C$=10, $n_T$=20}  & \multicolumn{5}{c|}{$n_C$=50, $n_T=100$} \\      \cmidrule{3-12}
\multicolumn{2}{c}{}  &  \multicolumn{10}{|c|}{$\delta$}  \\ 
\multicolumn{2}{c|}{} & 0 & 0.25 & 0.5 & 0.75 & 1  & 0 & 0.1 & 0.2 & 0.3 & 0.4 \\ \midrule
& BD & 2.49 & 12.15 & 35.37 & 66.11 & 88.61 & 2.49 & 10.54 & 29.31 & 56.60 & 80.75 \\ 
&  FD & 2.49 & 9.49 & 25.24 & 49.11 & 73.29 & 2.49 & 8.42 & 21.14 & 41.02 & 63.75 \\ 
&  RMD-Unit & 2.05 & 10.28 & 30.91 & 60.45 & 84.34 & 2.13 & 9.20 & 26.06 & 51.92 & 76.25 \\  \multirow{2}{*}{\parbox{3cm}{\centering Uncalibrated}} &  CDD & 2.26 & 10.64 & 31.07 & 59.87 & 83.59 & 2.26 & 9.28 & 25.67 & 50.68 & 74.79 \\ 
 & CDC  & 2.26 & 10.64 & 31.07 & 59.86 & 83.58 & 2.26 & 9.27 & 25.66 & 50.67 & 74.78 \\        \cmidrule{2-12}
&  EBPowD & 2.83 & 12.42 & 34.32 & 63.19 & 85.46 & 2.83 & 10.85 & 28.63 & 54.26 & 77.39 \\ 
    \multirow{2}{*}{\parbox{3cm}{\centering Calibrated to EBPowD}} &  CDD& 2.78 & 12.28 & 34.19 & 63.48 & 86.18 & 2.78 & 10.74 & 28.55 & 54.40 & 78.04 \\ 
  & CDC& 2.77 & 12.28 & 34.19 & 63.48 & 86.18 & 2.78 & 10.74 & 28.54 & 54.39 & 78.05 \\     \cmidrule{2-12}
    &  RMD-Unit (w=0.75)& 2.08 & 10.46 & 31.40 & 61.13 & 84.90 & 2.16 & 9.33 & 26.42 & 52.55 & 76.87 \\   
 \multirow{2}{*}{\parbox{3cm}{\centering Calibrated to RMD-Unit (w=0.75)}} &  CDD   & 2.21 & 10.86 & 32.18 & 61.95 & 85.48 & 2.26 & 9.58 & 26.96 & 53.17 & 77.48 \\ 
 & CDC & 2.21 & 10.84 & 32.14 & 61.91 & 85.46 & 2.25 & 9.56 & 26.91 & 53.13 & 77.46 \\         \cmidrule{2-12}
    & RMD-Unit (w=0.5) & 1.97 & 9.74 & 29.25 & 57.84 & 82.25 & 2.04 & 8.71 & 24.66 & 49.52 & 73.75 \\ 
 \multirow{2}{*}{\parbox{3cm}{\centering Calibrated to RMD-Unit (w=0.5)}}   &  CDD  & 2.02 & 9.95 & 29.85 & 58.64 & 82.77 & 2.10 & 8.93 & 25.10 & 50.15 & 74.45 \\ 
 & CDC  & 2.01 & 9.93 & 29.83 & 58.62 & 82.76 & 2.10 & 8.92 & 25.09 & 50.13 & 74.44 \\        \cmidrule{2-12}
    & RMD-Unit (w=0.25) & 1.98 & 9.34 & 27.49 & 54.75 & 79.53 & 2.04 & 8.26 & 22.88 & 46.11 & 70.14 \\   
  \multirow{2}{*}{\parbox{3cm}{\centering Calibrated to RMD-Unit (w=0.25)}}  & CDD    & 1.96 & 9.72 & 29.24 & 57.81 & 82.07 & 2.05 & 8.35 & 23.14 & 46.54 & 70.62 \\ 
 & CDC  & 1.96 & 9.70 & 29.22 & 57.79 & 82.06 & 2.05 & 8.35 & 23.14 & 46.54 & 70.62 \\
       \bottomrule
\end{tabular}
\caption{Average TIE rates (\%) and power (\%) for varying treatment effects $\delta$.  Normal outcomes with known variance $\sigma^2=1$, $n_C=10,n_T=20$ and $n_C=50,n_T=100$.  Averages are obtained by integrating over the informative prior for the control arm mean $\pi_C=N(0,\sigma_C=\sigma/\sqrt{n_{0C}})$, $n_{0C}=n_C$, corresponding to the analysis prior for the BD. $t = 4$ and $p = 4$ for the CDD. In the uncalibrated scenario, for the CDs $\alpha^{LOW}=0.01$  and $\alpha^{UP}=0.075$, and the weight of the mixture prior $w=0.7$. In each calibrated scenario, $\alpha^{LOW}$ and $\alpha^{UP}$ for the CDC and CDD are set to the minimum and maximum TIE rate observed over the conflict range [-2,2] for the competitor approach.}
\label{tab:avgOC3}
\end{table}

\end{document}